\documentclass[a4paper,11pt]{article}
\usepackage[a4paper,top=2cm,bottom=2cm,left=3cm,right=3cm,marginparwidth=1.75cm]{geometry}
\usepackage[pdftex, pdftitle={Article}, pdfauthor={Author}]{hyperref}
\usepackage[utf8]{inputenc}
\usepackage{float,array,multirow}
\usepackage{subcaption}
\usepackage{todonotes}
\usepackage{amsmath}
\usepackage{amssymb}
\usepackage[english]{babel}
\usepackage[dvipsnames]{xcolor}
\usepackage{pifont}
\usepackage{graphicx}
\usepackage{wrapfig}

\usepackage{physics}
\usepackage{placeins}
\usepackage{lscape}
\usepackage{rotfloat}
\usepackage{subcaption} 

\usepackage[backend=biber,maxnames=2,style=numeric-comp,sorting=none]{biblatex}
\DeclareFieldFormat[article]{pages}{#1}
\DeclareBibliographyAlias{article}{std}
\DeclareBibliographyAlias{online}{std}
\DeclareBibliographyAlias{book}{std}
\DeclareBibliographyDriver{std}{
  \usebibmacro{bibindex}
  \usebibmacro{begentry}
  \usebibmacro{author/editor+others/translator+others}
  \setunit{\labelnamepunct}\newblock
  \newunit\newblock
  \usebibmacro{date}
  \newunit\newblock
  \usebibmacro{journal}
  \newunit\newblock
  \printfield{volume}
  \newunit\newblock
  \printfield{pages}
  \newunit\newblock
  \usebibmacro{finentry}}

\begin{document}
\clearpage\thispagestyle{empty}
\vspace{3cm}
\begin{center}
\section*{Variability of MHD Instabilities in Benign Termination\\of High-Current Runaway Electron Beams\\in the JET and DIII-D Tokamaks}
\vspace{0.8cm}
C. F. B. Zimmermann\textsuperscript{1 *}, 
C. Paz-Soldan\textsuperscript{1}, 
G. Su\textsuperscript{1}, 
C. Reux\textsuperscript{2},\\ 
A. F. Battey\textsuperscript{1 \textdagger},
O. Ficker\textsuperscript{3},  
S. N. Gerasimov\textsuperscript{4},  
C. J. Hansen\textsuperscript{1}, 
S. Jachmich\textsuperscript{5},\\
A. Lvovskiy\textsuperscript{6},
J. Puchmayr\textsuperscript{7},
N. Schoonheere\textsuperscript{2}, 
U. Sheikh\textsuperscript{8}, 
I. G. Stewart\textsuperscript{1},  
G. Szepesi\textsuperscript{4},\\
JET Contributors\textsuperscript{9} and  
the EUROfusion Tokamak Exploitation Team\textsuperscript{10}\\
\vspace{0.6cm}
\textit{\textsuperscript{1} Department of Applied Physics and Applied Mathematics, Columbia University, New York 10027, USA}\\
\textit{\textsuperscript{2} CEA-IRFM, 13108 Saint-Paul-les-Durance, France}\\
\textit{\textsuperscript{3} Institute of Plasma Physics of the CAS, U Slovankou 1770/3, 182 00 Praha 8, CZ}\\
\textit{\textsuperscript{4} UKAEA, Culham Campus, Abingdon, Oxon, OX14 3DB, United Kingdom}\\
\textit{\textsuperscript{5} ITER Organization, 13108 Saint-Paul-les-Durance, France}\\
\textit{\textsuperscript{6} General Atomics, P.O. Box 85608 San Diego, California 92186, USA}\\
\textit{\textsuperscript{7} Max-Planck-Institut für Plasmaphysik, Boltzmannstr. 2, 85748 Garching, Germany}\\
\textit{\textsuperscript{8} Ecole Polytechnique Fédérale de Lausanne, SPC, CH-1015 Lausanne, Switzerland}\\
\textit{\textsuperscript{9} see the author list of C. F. Maggi et al. 2024 \textit{Nucl. Fusion} 64 112012}\\
\textit{\textsuperscript{10} see the author list of E. Joffrin et al. 2024 \textit{Nucl. Fusion} 64 112019}\\

\vspace{0.6cm}
\textsuperscript{*} E-mail of the corresponding author: benedikt.zimmermann@columbia.edu\\
\textsuperscript{\textdagger} current address: see \textit{\textsuperscript{8} }\\
\end{center}

\begin{abstract}
\noindent Benign termination, in which magnetohydrodynamic (MHD) instabilities deconfine runaway electrons (REs) following hydrogenic injections, is a promising strategy for mitigating dangerous RE loads after disruptions. Recent experiments on the Joint European Torus (JET) have explored this scenario at higher pre-disruptive plasma currents than are achievable on other devices, revealing challenges in obtaining benign terminations at $I_p \geq 2.5$ MA. This work analyzes the evolution of these high-current RE beams and their terminating MHD events using fast magnetic sensor measurements and EFIT equilibrium reconstructions for approximately $40$ JET and $20$ DIII-D tokamak discharges. On JET, unsuccessful non-benign terminations occur at low edge safety factor ($q_{\text{edge}} \approx 2$), and are preceded by intermittent, non-terminating MHD events at higher rational $q_{\text{edge}}$. Trends in the internal inductance $l_i$ indicate more peaked RE current profiles in the high-$I_p$ non-benign population, which may hinder successful recombination through re-ionization of the companion plasma. In contrast, benign terminations on JET typically occur at higher $q_{\text{edge}} \geq 3$ and exhibit less peaked RE current profiles. DIII-D displays a broader range of terminating edge safety factors, again correlated with the measured $l_i$ values. Across both tokamaks, the RE current peaking is therefore found to determine which MHD instability boundary is encountered, a result confirmed by linear resistive MHD modeling with the CASTOR3D code. Measured growth rates are similar for benign and non-benign cases, indicating that ideal MHD timescales at low density after hydrogenic injection do not alone explain efficient RE deconfinement. Instead, non-benign cases are most readily characterized by their comparably lower overall MHD perturbation amplitudes $\delta B$. These observations suggest that the interplay between ideal and resistive dynamics governs the termination process, with implications for extrapolating benign RE termination to high-$I_p$ reactor scenarios.
\end{abstract}
\newpage

\section{Introduction}

Mitigation schemes for plasma disruptions will be essential for future high-current tokamaks such as ITER or commercial fusion power plants, as they can impose significant thermal and mechanical loads on tokamak components \cite{Loarte_2011}. A promising solution to mitigate the resulting runaway electron (RE) beams is the \textit{benign termination} scheme, in which hydrogen is injected into the RE beam and a magnetohydrodynamic (MHD) instability deconfines the REs, increases the RE deposition wetted area, and terminates the RE beam \textit{benignly}. After being first observed on DIII-D \cite{pazsoldan2019ppcf}, dedicated experiments were performed on the Joint European Torus (JET) tokamak \cite{reux2021prl,reux2022ppcf}, along with the successful elaboration of this technique on DIII-D. This led to modeling work by Liu \textit{et al.} in Ref.~\cite{liu2019nf} and to a publication by Paz-Soldan \textit{et al.} \cite{pazsoldan2021nf}, which included further MHD analysis of the DIII-D data as well as a comparison with the JET data published by Reux \textit{et al.} In addition, Sheikh \textit{et al.} contributed a study on the TCV and AUG tokamaks \cite{sheikh2024ppcf}, focusing on the operational windows for benign termination and the observed phenomenology with a brief discussion of measured MHD growth rates. In all these experiments, it was observed that there are limits to the amount of injected material for which the injection recombines the background, or \textit{companion}, plasma. While on DIII-D this recombination was observed as a key component in making the mitigation both successful and benign, the corresponding density measurements on JET are not available for a number of discharges due to compromised diagnostic data and insufficient spatial resolution of the lines of sight of the interferometry systems. In addition, the interpretation of the pressure gauge signals was found to be hampered in these discharges \cite{Schoonheere_2024}. However, all available data suggest that, also on JET, a successful recombination of the background plasma is the key requirement for a benign termination \cite{reux2024runaway}.\par 
First publications on the effect of material injections into RE beams in JET by Reux \textit{et al.} used data collected before 2022 and, among other things, investigated the termination and mitigation characteristics in relation to the secondary injection material employed \cite{reux2021prl,reux2022ppcf}. Not all of these early experiments were aimed at achieving a benign termination as defined by Paz-Soldan \textit{et al} \cite{pazsoldan2019ppcf}. In general, it was found that high-$Z$ or mixed secondary injections predominantly led to unsuccessful RE terminations. For such terminations, particularly with impurity fractions above 5\%, a rich phenomenology was observed due to the variety of kinetic processes involved. Paz-Soldan \textit{et al.} revisited the JET dataset and compared it with similar data from DIII-D \cite{pazsoldan2021nf}. That work placed greater emphasis on investigating the MHD and equilibrium properties, although discharges with various secondary injection mixtures were also presented. At the time, it was concluded that successful and safe benign terminations were more likely to occur in the absence of high-$Z$ material. This finding motivated further JET experiments focusing on clean hydrogenic injections. The present publication differs from the work in Refs. \cite{pazsoldan2021nf,reux2022ppcf} by restricting itself to the subset of JET and DIII-D discharges with hydrogenic injections and impurity fractions below 5\%, which have a higher fraction of successful, benign terminations. However, low impurity fractions are not a sufficient criteria for a benign termination characteristic.\par

\begin{figure}
    \centering
    \includegraphics[width=0.5\linewidth, trim=0 0 0 0, clip]{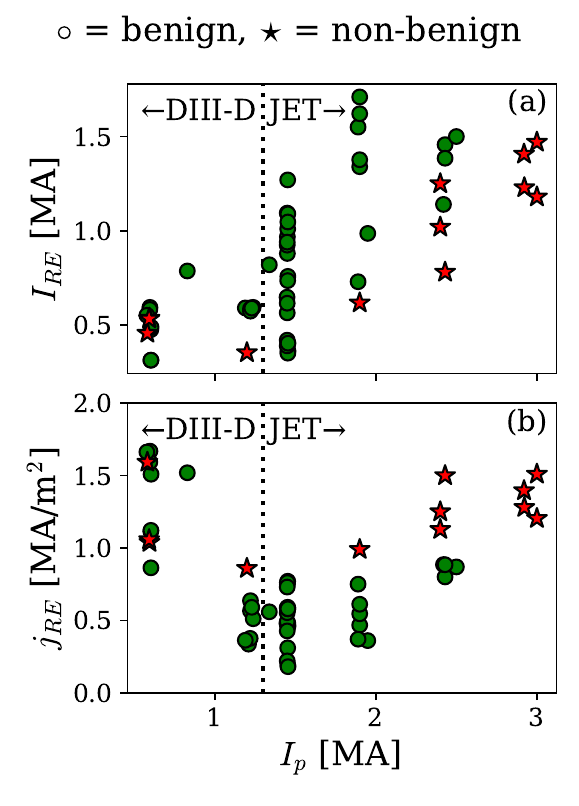}
    \caption{The studied database for DIII-D (l.h.s.) and JET (r.h.s.). Panel (a): Runaway current \(I_{RE}\) immediately before the terminating MHD event versus the pre-disruptive plasma current \(I_p\). Panel (b): Runaway current density \(j_{RE}\) immediately before the terminating MHD event versus the pre-disruptive plasma current \(I_p\). On JET, non-benign cases terminate from higher RE current densities. DIII-D does not show a separation of benign and non-benign cases by the RE current density at low pre-disruptive $I_p$, despite reaching similar RE current densities. Benign termination cases are marked with a green circle; non-benign cases with a red star.} 
    \label{fig:JET_Ip_IRE_jRE}
\end{figure}

Furthermore, all the JET results published so far in Refs. \cite{pazsoldan2021nf,reux2022ppcf} were limited to discharges with pre-disruptive plasma currents below $1.5$~MA (JET discharge numbers below $100,000$). Recent experiments on JET in 2022 and 2023 have reached higher pre-disruptive plasma and RE currents than on other machines, up to $I_p \approx 3$ MA. They have revealed challenges in achieving benign termination at high pre-disruptive current values of $I_p \geq 2.5$~MA. This is illustrated in Fig.~\ref{fig:JET_Ip_IRE_jRE}(a), which shows the pre-disruptive plasma current $I_p$ on the x-axis and the RE current prior to termination on the y-axis, with only non-benign, red cases for $I_p \approx 3.0$~MA on JET (r.h.s.). All cases with $I_p > 1.5$~MA stem from JET campaigns in 2022 and 2023 and have not yet been published in peer-reviewed literature, though a short description can be found in Ref. \cite{reux2024runaway}. In general, understanding the distinguishing features of these high-$I_p$ experiments on JET is critical for reactor-relevant extrapolation. In particular, studying the dynamics of the terminating MHD activity is crucial, as a conclusive explanation for what differentiates the nature of the terminating MHD event between benign and non-benign cases is still lacking. In addition, the referenced publications on DIII-D and JET, in particular by Paz-Soldan \textit{et al.} \cite{pazsoldan2021nf}, highlighted a phenomenological difference with benign termination on DIII-D at the lowest edge safety factors (reproduced on AUG and TCV \cite{sheikh2024ppcf}), while on JET, benign terminations at higher edge safety factors were also observed. This poses an open question of how observations from JET compare to those on DIII-D, TCV, and AUG, and how to unify the observed phenomenology.\par
The present analysis addresses these questions through a systematic and empirical study of magnetic sensor data from a database of about 40 RE discharges collected during the JET campaigns between 2019–2023 that explored the benign termination scenario and its limits and compares them with about 20 RE discharges from DIII-D from 2018, 2021, and 2022. For both datasets, only discharges with pure hydrogenic injections are analyzed. This paper is structured as follows. Section~\ref{sec:scenario_method} describes the experimental scenario, datasets, diagnostics, and analysis methods. Section~\ref{sec:analysis} examines the RE beam evolution on JET prior to termination (Sec.~\ref{sec:evolution_high_Ip}), documents intermittent MHD activity preceding the final termination event on JET (Sec.~\ref{sec:intermittent_events_JET}), and compares the phenomenology of the terminating instabilities on JET and DIII-D (Sec.~\ref{sec:phenomenology_termination}). Section~\ref{sec:modeling} outlines dedicated MHD modeling and discusses MHD stability boundaries relevant to the observed terminations. Section~\ref{sec:summary_outlook} summarizes the main findings and discusses implications and priorities for future work.

%
\section{Scenario and Methodology}
\label{sec:scenario_method}
%
\begin{figure}[htbp]
    \centering
    \begin{minipage}{0.48\textwidth}
        \centering
        Benign termination on JET \#102617\par\medskip
        \includegraphics[width=\linewidth]{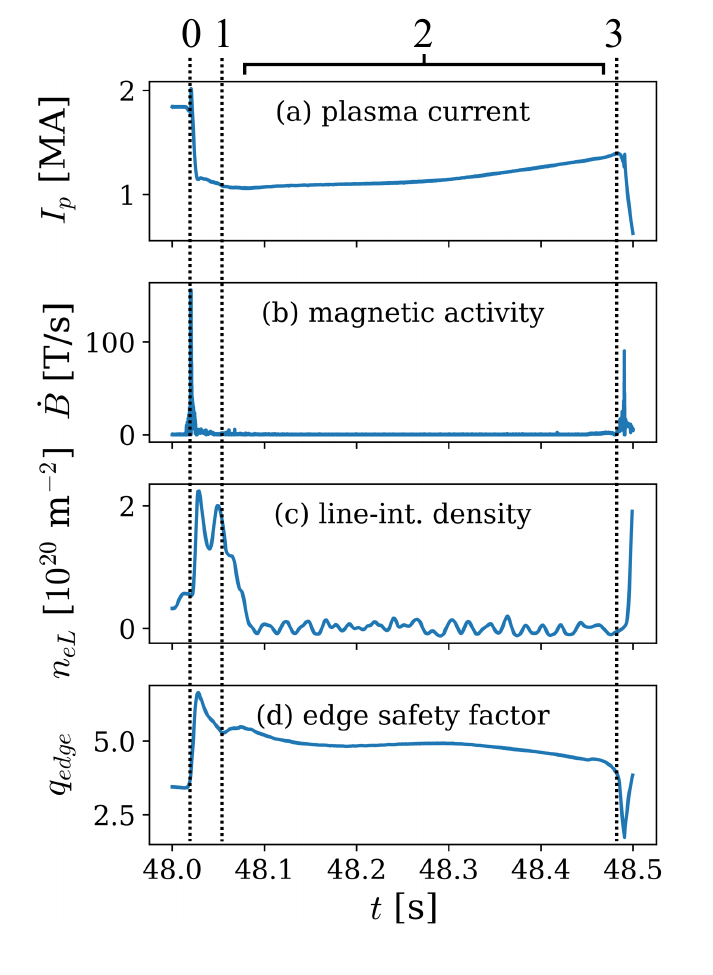}
        \captionof{figure}{Typical benign termination on JET (\#102617), marking the primary disruption (0), the hydrogenic injection (1), the RE plateau phase with plasma compression towards the center post (2), and the onset of the terminating MHD event (3).}
        \label{fig:overview_timetrace}
    \end{minipage}
    \hfill
    \begin{minipage}{0.48\textwidth}
        \centering
        Non-benign termination on JET \#105794\par\medskip
        \includegraphics[width=\linewidth]{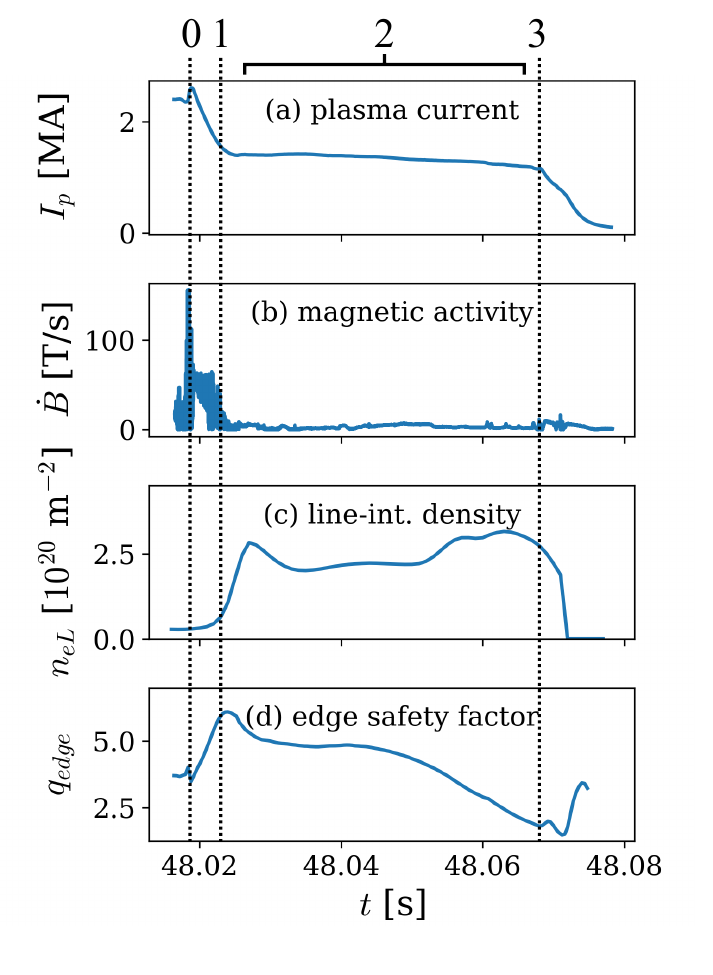}
        \captionof{figure}{Typical non-benign termination on JET (\#105794), marking the primary disruption (0), the hydrogenic injection (1), the RE shorter plateau phase with plasma compression towards the center post (2), and the onset of the termination (3), without significant magnetic activity.}
        \label{fig:overview_timetrace_nonbenign}
    \end{minipage}
\end{figure}

Typical benign termination experiments feature a short pre-disruptive flattop phase to ramp up the plasma current to the desired values. At this stage, significant energy is stored in the poloidal magnetic fields generated by the plasma current. To study a post-disruptive RE beam, a primary disruption is then (artificially) triggered by means of high-$Z$ gas puffing or shattered pellet injection (SPI), e.g., through argon injection. This event is marked by the left vertical dotted line (0) in Fig.~\ref{fig:overview_timetrace}. During this primary disruption, the plasma current, shown in Panel (a), drops, and strong signatures appear in the magnetic measurements, here illustrated by the signal from a single Mirnov coil in Panel (b). After the thermal quench of the primary disruption, the hot-tail \cite{Chiu_1998}, avalanche \cite{Rosenbluth_1997}, and Dreicer mechanisms \cite{Dreicer_1959} generate a relativistic runaway current, such that after several tens of milliseconds the entire plasma current is carried by REs. For a more detailed discussion of these processes, as well as a general introduction to RE physics, the reader is referred to the excellent review by Breizman \textit{et al.} \cite{Breizman_2019}.\par

Up to this point, this procedure is typical of a generic RE experiment. To now attempt a benign termination, a secondary, hydrogenic injection is induced via SPI, massive gas injection, or the disruption mitigation valves, see vertical dotted line (1). This work focuses on clean hydrogen (deuterium or protium) secondary injections. Depending on the underlying kinetics, as discussed for example by Hollmann \textit{et al.} \cite{Hollmann2023} and Hoppe \textit{et al.} \cite{Hoppe2025}, the injected material either recombines the companion plasma, typically leading to a benign termination, or remains ionized, in which case a non-benign termination usually follows. Panel (c) depicts the line-integrated density, measured with interferometry, for a successful recombination of the companion plasma with a drop of the measurement signal below noise level after the secondary injection.\par 
Consequently, the RE beam is deliberately driven MHD-unstable, either by reducing the edge safety factor to low rational values through compression of the plasma column against the center post, or by elongating the beam, switching off the control system, and provoking a vertical displacement event (VDE). This phase is indicated by (2) in Fig.~\ref{fig:overview_timetrace}. An overview of the variety of MHD dynamics involved can be found, for example, in the textbooks by Freidberg \cite{freidberg1982ideal}, Zohm \cite{zohm2015magnetohydrodynamic}, and Igochine \textit{et al.} \cite{igochine2015active}. The most basic MHD mode responsible for the majority of events discussed in this work is the so-called \emph{kink mode}.\par 
Both approaches of driving the beam unstable can trigger a large-scale MHD event and terminate the RE beam either without (benign) or with (non-benign) localized impact of the RE beam on the first wall or divertor structure. This is observed after the onset of the termination (3), characterized by a drop in plasma current and safety factor, together with a spike in magnetic activity. In this work, most of the data points, if not otherwise explicitly mentioned, are sampled just prior to the termination. Specifically, this refers to the time point of the last equilibrium calculation before the magnetic signature rises, marking the onset of the terminating MHD event. On JET, the classification of benign and non-benign terminations is based on IR camera observations and the corresponding heat fluxes, as well as on the scheme developed by Loarte \textit{et al.} \cite{Loarte_2011}, which tracks the conversion of magnetic energy into kinetic energy, a process that is more efficient in non-benign cases. Both is documented in Reux \textit{et al.} \cite{reux2022ppcf}. On DIII-D, due to the lack of comparable IR camera coverage, measured hard X-ray (HXR) spikes serve as a indicator distinguishing clean, benign terminations from non-benign terminations in which the RE beam strikes the wall, emitting HXR, as already applied in \cite{pazsoldan2021nf}. It seems important to highlight that the community has not yet settled on a unified classification scheme for the characteristics of terminations and that, for consistency with previous work, this study adopts the respective definitions used on each machine. However, Paz-Soldan \textit{et al.} verified the agreement between the HXR and IR classification schemes for selected experiments on DIII-D, as discussed in more detail in \cite{pazsoldan2021nf}. A more unified classification scheme should be pursued in the future, but this is beyond the scope of this work.\par
%
%
A typical non-benign termination is shown in Fig.~\ref{fig:overview_timetrace_nonbenign}. As in the benign counterpart, Panel (a) shows the evolution of the plasma current. The entire RE phase is significantly shorter than in the benign case. The final loss event, see (3), is not abrupt but rather a slow decay of the REs, which are not efficiently deconfined. This can be seen in Panel (b), in which the magnetic activity during the termination is insignificant, indicating an inefficient MHD event in the non-benign case.  The line-integrated density in Panel (c) suggests that the recombination of the free electrons was not successful, the material injection leads to a large increase of the measurement after (1). The signal drops steeply to zero at the termination when the residual plasma is lost and deconfined. The edge safety factor shown in Panel (d) illustrates the rapid compression of the RE beam and indicates that a low edge safety factor is reached prior to the onset of the termination at (3).\par
\begin{figure}
    \centering
    \includegraphics[width=0.45\linewidth]{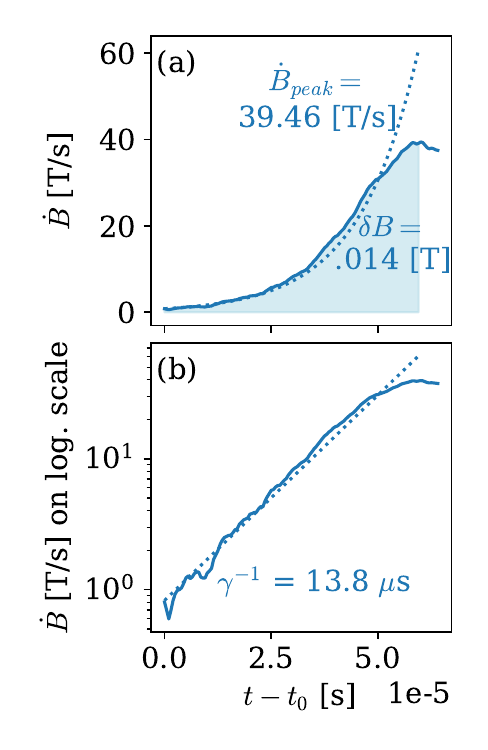}
    \caption{Magnetic analysis of the terminating MHD event in the JET discharge \#102617. Panel (a) shows the time trace of $\dot{B}$ of the $n=1$ component reconstructed from the toroidal Mirnov coil array (solid line), with the shaded area indicating the integrated perturbation amplitude $\delta B$ over the time span shown in the plot. A fitted exponential growth is overlaid (dotted line) for comparison. Panel (b) displays $\dot{B}$ on a logarithmic scale (solid line) together with a linear fit (dotted line), from which the growth rate $\gamma$ is extracted.}
    \label{fig:magnetics_analysis}
\end{figure}
A more detailed investigation of the MHD events in this work is based on fast magnetic measurements for MHD mode decomposition and equilibrium reconstruction. The primary diagnostics for the magnetic measurements are the Mirnov coils, which are small pickup coils placed around the tokamak vessel to measure time-varying magnetic fields. They operate on the principle of electromagnetic induction: a changing magnetic flux through the coil induces a voltage proportional to $\dot{B}$, the time derivative of the perturbed magnetic field. An example of such a measurement during the terminating MHD event of the JET discharge \#102617 is shown in Panel (a) of Fig.~\ref{fig:magnetics_analysis} (solid line). Arrays of Mirnov coils distributed toroidally enable the mode decomposition of the measured perturbations in terms of the relative phases and amplitudes for different mode numbers $n$ through matrix operations \cite{Munaretto_2021}. While DIII-D provides a large number of magnetic arrays, at JET only $5$–$7$ fast Mirnov coils were usable, depending on the plasma discharge. This restricts the analysis to low mode numbers $n=1,2$. Because of their fast response and sampling rates on the MHz scale, Mirnov coils are well suited for studying MHD instabilities, mode growth, and termination dynamics, where rapid changes in the magnetic field occur. Integration of the coil signals over the MHD event correlated to the final loss event yields $\delta B$, providing direct information on the amplitude of the underlying magnetic perturbations, as indicated by the shaded region under the curve in Panel (a). By fitting a linear curve, dotted line in Panel (b), to the logarithmic $\dot{B}$ values (solid line), the growth rate $\gamma$ can be measured. Fitting $\gamma$ with this approach is valid only during the initial exponential growth phase, when nonlinear coupling and saturation are not yet occurring. In this regime, fitting $\dot{B}$ or $\delta B$ yields identical results, differing only by constant offsets in the linear fit. The same constant offsets are introduced when considering conducting and shielding wall structures in front of magnetic sensors and, therefore, do not affect the results presented here. It should be noted that the Mirnov coil arrays used in this work measure the perturbed poloidal magnetic field $\delta B_p$, which, for simplicity, is referred to as $\delta B$ throughout this work. During the course of this work, the use of a normalization to the distance between the measurement array and the last closed magnetic flux surface was attempted, and the resulting $\delta B$ values were carefully compared. The normalized and un-normalized $\delta B$ values were found to exhibit an approximate 1:1 relationship, with some unsystematic scatter. The distance used for normalization averaged $1.04$ m (with a standard deviation of $0.35$ m) and was therefore not found to significantly alter the trends observed in the subsequent analysis. A consistent normalization would require knowledge of the RE beam position during the termination events, as well as information about the poloidal mode structure \cite{igochine2015active}, neither of which can be obtained in the studied scenario. Therefore, the authors refrain from applying any normalization. As a final caveat, it should be noted that significant efforts were made during the course of this project to automate the identification of the time window for the integration of $\delta B$ from the $\dot{B}$ signal, as well as the fitting window for the exponential growth rates. However, due to the large variety of mode characteristics encountered, such as non-linear damping, mode rotation, and $n=0$ pollution, a very careful manual identification of the integration and fitting time windows was ultimately required. This process relied on identifying the first strong $\dot{B}$ peak correlated with RE deconfinement and integrating from the associated rise of the left flank out of the noise to its decline. Establishing automated workflows for large database studies would require dedicated analysis projects and is beyond the scope of this work.\par
Equilibrium reconstruction is a method used to infer the spatial distribution of plasma current and pressure from external magnetic measurements by solving the underlying MHD equilibrium equations. In this work, the EFIT++ (for JET) and the EFIT codes (for DIII-D) \cite{Lao_1985} solve the Grad-Shafranov equation \cite{GradRubin1958,Shafranov1966} iteratively, adjusting model parameters until the calculated magnetic signals match the experimental data from magnetic probes, flux loops, and diamagnetic measurements. This procedure yields a self-consistent two-dimensional equilibrium, providing the time-resolved evolution of the plasma boundary, elongation $\kappa$, cross-section $A$, safety factor profile $q$, and internal current distribution, which is related to the internal inductance $l_i$. In this work, the $l_i(3)$ definition is used. For disruptive scenarios, high temporal resolution and careful convergence settings are required, since rapid changes in current profiles and plasma position can challenge standard equilibrium solvers. The equilibria analyzed in this study were therefore specifically generated to investigate disruption properties. It should be noted that, for both machines, the reconstruction of the current profiles based on soft X-ray or bremsstrahlung measurements (as done for example in Refs. \cite{Loarte_2011,marini2024runaway}) was unsuccessful, as the majority of the plasma discharges either had already recombined (benign, purged, with no background impurities, and no emissions), were too transient, or the corresponding diagnostic data is not usable. However, the work by Loarte \textit{et al.} demonstrates that, for example, the internal inductance of RE beams reconstructed from X-rays, when compared with EFIT, shows no systematic deviation but rather a scattering \cite[Fig.~6]{Loarte_2011}. Relativistic corrections are not applied in the equilibrium reconstruction and may represent a source of error that cannot be accounted for.
%
\section{Experimental Analysis}
\label{sec:analysis}

The following section presents a detailed examination of experimental observations from JET and DIII-D concerning the evolution and termination of RE beams. Emphasis is placed on identifying and understanding characteristic differences in the observed phenomenology of the equilibrium evolution and MHD behavior between benign and non-benign terminations. To elucidate the associated MHD behavior, particular attention is given to the roles of current peaking and the edge safety factor.

\subsection{Evolution of high-$I_p$ RE beams on JET}
\label{sec:evolution_high_Ip}
\begin{figure}
    \centering
    \includegraphics[width=0.66\textwidth, trim=0 0 0 0, clip]{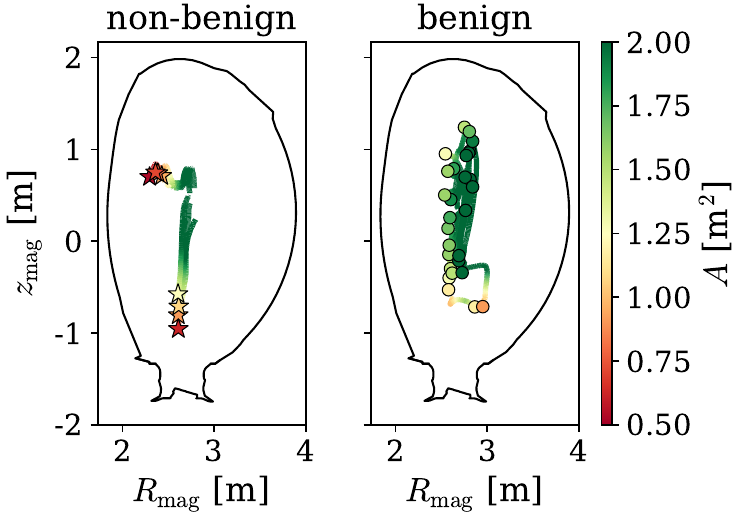}
    \caption{Trajectory of the RE beam center before the onset of the terminating MHD event in JET. Color coding is based on the plasma cross-section. The JET wall contour is shown in black. Shown are several trajectories for different discharges in the studied database.}
    \label{fig:Rz_trajectory}
\end{figure}

The first critical observation concerns the fundamental difference in the duration of the RE plateau phase between benign and non-benign cases. In typical benign scenarios, the RE plateau phase is long, controlled, and stable, with an average duration of $0.62$~s (standard deviation $0.66$~s). In contrast, in cases where recombination is unsuccessful and the outcome is non-benign, no stable plateau forms, with an average duration of $0.07$~s (standard deviation $0.03$~s) before the termination. This is likely due to the higher resistivity of the non-recombined plasma, which results in a situation where the spatial position of the RE beam is no longer effectively controlled. The subsequent MHD response, however, is assumed to be largely indifferent to the specific pathway by which the system becomes unstable. In addition, typical MHD time scales, as discussed later, are on the order of microseconds, such that this deviation in RE plateau length cannot systematically interact with the time the MHD instabilities have to develop. A more detailed investigation of how the RE trajectory influences key quantities, such as the current profile distribution, and, consequently, the MHD instability is left for future work and cannot be disentangled from present experiments.\par 

The next step of the experimental analysis concerns the properties of the plasma equilibrium. In Fig.~\ref{fig:Rz_trajectory}, the trajectory of the reconstructed magnetic axis is plotted for the last 300~ms before termination in the $R$-$z$-space with the JET contour in black for comparison. For cases in which the RE plateau phase was shorter, sampling was performed from the onset of the RE plateau phase. The panel on the left shows the non-benign cases, while the panel on the right shows the benign cases. The color coding corresponds to the cross-section of the RE column. It is observed that the onset of termination in non-benign cases occurs at a smaller cross-section and in a less central position, whereas benign terminations typically begin from a relatively larger cross-section and a more central position. This suggests a different evolution of the RE beam in benign and non-benign cases, with non-benign beams being compressed to smaller cross-sections and, consequently, to lower edge safety factors than their benign counterparts. As already noted, the non-benign cases on JET are usually very short, suggesting that the higher resistivity of non-recombined plasmas likely affects their controllability. Under these circumstances, the short lifetime of the non-recombined plasmas could interact with the evolution of the current profiles. Assuming identical current-profile evolution across shots, for example through competing time scales of different RE generation or current diffusion and redistribution mechanisms, any systematic variation in duration would alter the current profile at the point of the secondary disruption. A realistic modeling of the RE evolution and formation, as already stated, is the subject of a subsequent publication on this topic.\par 
As discussed above, the non-benign cases on JET are associated with high pre-disruptive plasma currents. The higher stored energy in these cases leads to strong RE currents. Combined with the observed smaller cross-sections, this results in higher RE current densities prior to termination. Consequently, a clear separation between benign and non-benign cases emerges in terms of their RE current density $j_{RE}=I_{RE}/A$, as illustrated in Fig.~\ref{fig:JET_Ip_IRE_jRE}(b) with the JET datapoints on the r.h.s. This behavior is not commonly observed across the previously discussed devices, and a similar analysis for the studied DIII-D dataset does not show such a separation, see l.h.s. of the same figure. This suggests that different mechanisms play a role in the formation or termination of the RE currents on the two machines. Interestingly, at the highest \(I_p \approx 1.2\) MA on DIII-D, the only non-benign case does separate from the benign cases with thresholds similar to those in the JET dataset, but more data would be needed to draw firm conclusions. For JET, however, the higher RE current densities may interact with kinetic effects, as suggested by Hollmann and Hoppe, who noted that more concentrated (peaked) RE current density profiles may hinder effective recombination of the companion plasma or can even lead to re-ionization of an initially recombined background plasma \cite{Hollmann2023,Hoppe2025}.\par 
%
\subsection{Intermittent MHD events prior to the termination in JET}
\label{sec:intermittent_events_JET}

\begin{figure}
    \centering
    \includegraphics[width=0.66\textwidth, trim=0 0 0 0, clip]{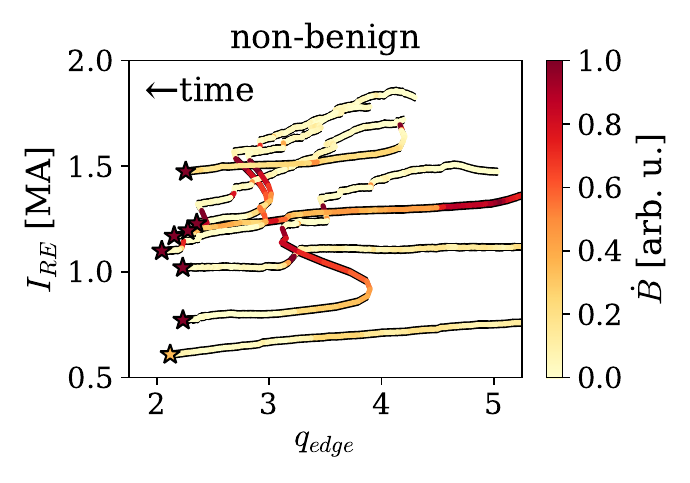}
    \caption{Trajectory of $I_{RE}$ versus $q_{\text{edge}}$ before the onset of the final current decay, shown for the non-benign cases in JET only. Coloring corresponds to $\dot{B}$ of a single probe to depict the MHD signature, showing non-terminating events at higher rational $q_{\text{edge}}$.}
    \label{fig:qedge_vs_IRE trajectories}
\end{figure}
As observed in the previous section, high-$I_p$, non-benign cases on JET tend to terminate from smaller plasma cross-sections. Figure~\ref{fig:qedge_vs_IRE trajectories} shows the trajectories of these non-benign cases before the onset of the final termination. It can be seen that the RE beams approach termination from higher edge safety factor values, with the time evolution proceeding from right to left. The color coding of the time traces indicates the magnetic activity measured by a single Mirnov coil. Most of the non-benign discharges undergo MHD events when passing rational values, e.g., at $q_{\text{edge}}=3,4$. While these events expel a fraction of the RE current in some cases, they are not capable of fully deconfining the RE beam before it reaches $q_{\text{edge}} \approx 2$. As discussed later in more detail, benign cases terminate earlier, latest at $q_{\text{edge}}=3$, and at larger plasma cross-sections. It should be noted that the trend observed in Fig.~\ref{fig:JET_Ip_IRE_jRE}(a), namely, that the non-benign cases with the highest $I_p$ values do not necessarily exhibit the largest $I_{RE}$ immediately prior to the final termination, correlates with the occurrence of the intermittent events, as well as with the rather short duration of the non-benign terminations, such that the RE beam does not fully form, as discussed earlier.\par 
\begin{figure}
    \centering
    \includegraphics[width=1\textwidth, trim=0 0 0 0, clip]{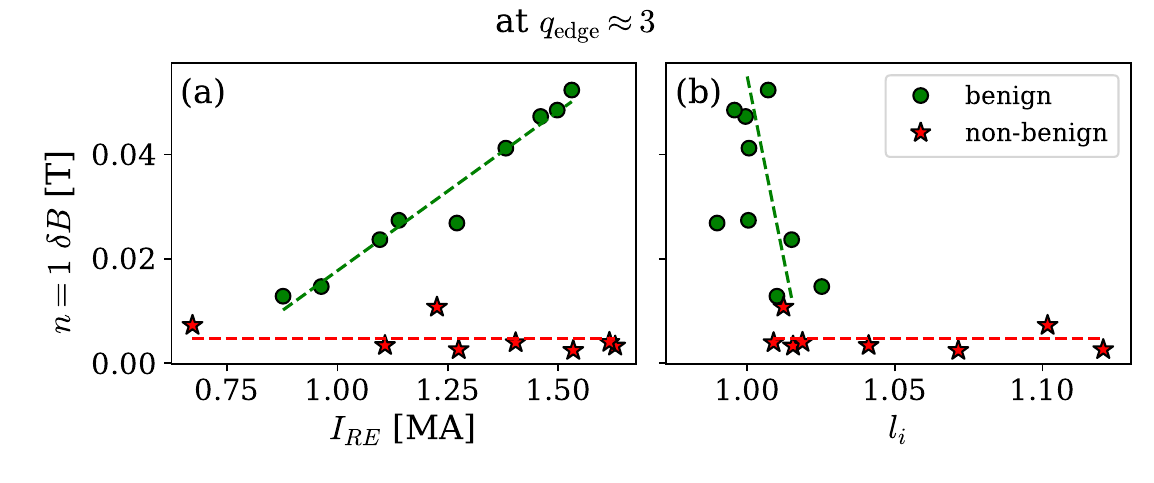}
    \caption{Perturbed magnetic field $\delta B$ integrated over the MHD events in $q_{\text{edge}} \approx 3$ for the benign (green circles) and non-benign cases (red stars) over the RE current, see Panel (a), and internal inductance, see Panel (b), for the JET dataset. One can see in both Panels that the non-benign cases have a very low MHD signature, while it increases linearly with the RE current for the benign cases. As shown in Panel (b), most of the benign cases have lower internal inductance, suggesting a less peaked RE current distribution.}
    \label{fig:deltaB_in_qedge3}
\end{figure}
In Fig.~\ref{fig:deltaB_in_qedge3}(a), the measured perturbed $\delta B$ is shown for benign and non-benign cases that pass through or terminate at $q_{\text{edge}} = 3$. The MHD amplitude in benign cases increases linearly with the RE current deconfined during the event. This suggests that there exists a threshold in $\delta B/B$ at which the MHD mode becomes strong enough to trigger a catastrophic, terminating MHD event. Here, $B$ represents the poloidal magnetic field, with $B_p \sim I_{RE}$ since $\delta B = \delta B_p$ and the variation in the toroidal magnetic field $B_t$ is small in this dataset, ranging only from $3.0$ to $3.45$ T. Non-benign cases exhibit a clearly suppressed MHD amplitude, insufficient to cause a proper termination. At the moment, it remains an open question how the recombination state of the companion plasma influences the amplitude or saturation of these MHD events.\par 
A possible explanation for the observed suppression is provided in Panel (b), which shows a tendency toward higher $l_i$ values in most of the non-benign cases, suggesting a more peaked RE current distribution. The intermittent small events do not change the measured $l_i$ values in the final phase prior to termination significantly. Variations in the current profiles could originate from the dependence of the RE current density distribution on the specific mechanisms by which the RE current forms during the primary disruption. JET, in particular, is characterized by strong RE avalanche generation, which could explain the different $j_{RE}$ profiles observed. This was discussed in detail by Smith \textit{et al.} \cite{Smith_2010} who predicted that the absolute pre-disruptive current can influence the RE current distribution on JET. Steepening of the RE current profiles, correlated with higher pre-disruptive $I_p$, was also observed on JET by Loarte \textit{et al.}~\cite{Loarte_2011}.\par 
Overall, these observations indicate that internal differences within the RE beam may govern the distinct MHD behavior of benign and non-benign cases in the high-$I_p$ dataset on JET. They suggest that the suppressed MHD activity in non-benign cases arises from a stronger peaking of the current profiles. It should be noted that the different localization prior to the final terminating MHD event, as shown in Fig.~\ref{fig:Rz_trajectory}, is more likely a result of subsequent compression (after an unsuccessful termination in $q_{\text{edge}} = 3$) rather than the initial cause of the differing evolution, as the non-benign cases are still rather central when they pass through $q_{\text{edge}} = 3$.
\subsection{Phenomenology of the terminating MHD events on JET and DIII-D}
\label{sec:phenomenology_termination}
\begin{figure}
    \centering
    \includegraphics[width=1\linewidth, trim=0 0 80 0, clip]{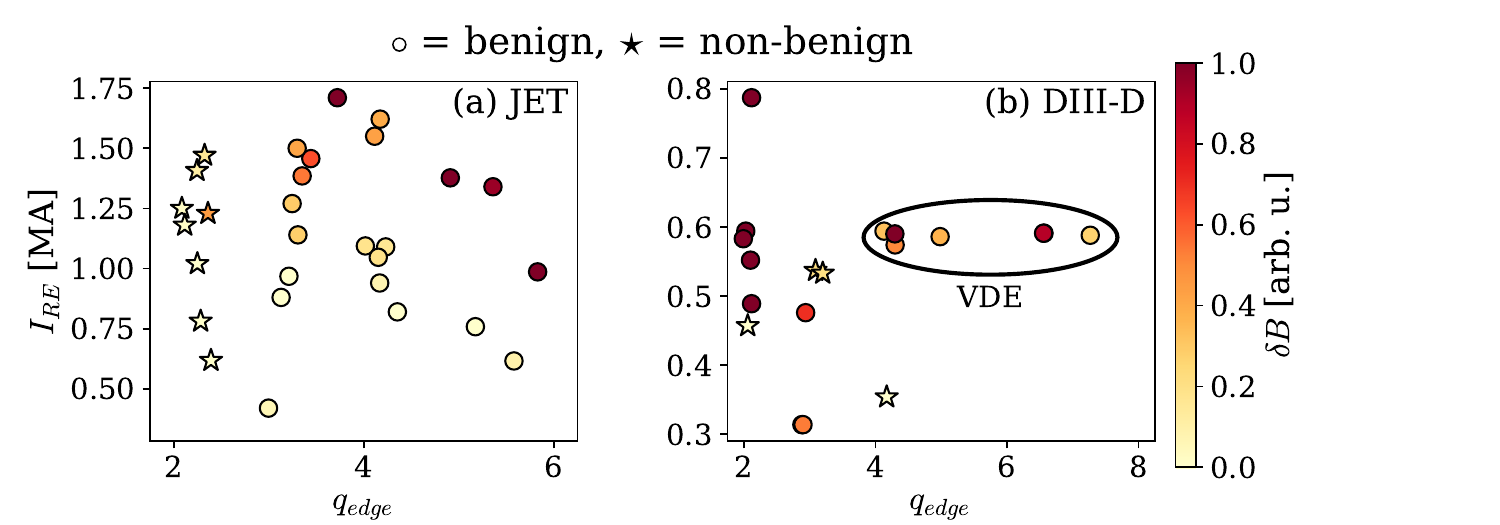}
    \caption{Runaway current $I_{RE}$ versus the edge safety factor $q_\text{edge}$ immediately before the terminating MHD event on JET (Panel a) and DIII-D (Panel b). Benign termination cases are marked with a circle, non-benign cases with a star. Color coding reflects the measured $n=1$ component of $\delta B$ during the terminating event, with normalization applied to facilitate comparison among the datasets. Aside from the encircled and marked cases on DIII-D, which disrupted via VDEs, all other cases relied on center-post compression.}
    \label{fig:IRE_vs_qedge_both_machines}
\end{figure}
The comparison of benign and non-benign cases in JET at $q_{\text{edge}} = 3$ in the previous section indicated that changes in current peaking could be responsible for the termination behavior. Expanding this analysis to the entire available database supports this hypothesis and reveals differences between JET and DIII-D.\par

Figure~\ref{fig:IRE_vs_qedge_both_machines} shows the RE current prior to the final termination on the y-axis and the measured $q_{\text{edge}}$ values on the x-axis, with JET results in Panel (a) and DIII-D results in Panel (b). For terminations with center-post compression, $q_{\text{edge}}$ corresponds to $q_{\text{limiter}}$, while for cases with VDEs, which are still diverted during the RE beam phase, it corresponds to $q_{95}$. It should be noted that, due to faulty data from single magnetic probes, the JET dataset for which a proper mode decomposition was possible is reduced from 37 cases in Fig.~\ref{fig:JET_Ip_IRE_jRE} to 29. Therefore, all subsequent discussions of the MHD properties refer to a slightly reduced dataset.\par  
In the JET dataset, all non-benign cases terminate at low $q_{\text{edge}} \approx 2$, whereas benign cases terminate at higher edge safety factors. On DIII-D, cases with VDEs terminate at high safety factors, while most benign cases with center-post compression terminate at low edge safety factor $q_{\text{edge}} \approx 2$. In contrast, three out of four non-benign cases on DIII-D terminate at slightly higher $q_{\text{edge}}$. The colorbar reflects the measured $\delta B$. A comparison of $\delta B$ across all cases shows that non-benign terminations are consistently associated with low amplitudes, whereas in both machines the MHD amplitude tends to be stronger to deconfine the RE beam benignly, particularly at higher RE currents. Normalizing the measured $\delta B$ to the distance between the magnetic sensors and the reconstructed last closed flux surface must be regarded with caution, as it relies on equilibrium reconstruction, which becomes inaccessible during the highly transient final disruption phase. Applying this normalization based on the equilibrium information prior to the termination does not significantly alter the conclusions discussed in this paper. However, it should be noted that the four points with the highest $\delta B$ values shown in Panel~(a) for JET correspond to cases that begin their termination closest to the magnetic sensors. This explains the proximity of some of the orange and deep red points.\par 
This work builds upon the DIII-D dataset studied in an earlier work by Paz-Soldan \textit{et al.}~\cite{pazsoldan2021nf}. That study considered a slightly different set of discharges with a wide mix of injection species and reported a phenomenology in which non-benign cases terminated at higher $q_{\text{edge}}$. In the present work, the dataset contains only hydrogenic injections and additionally includes cases with VDEs.\par
\begin{figure}
    \centering
    \includegraphics[width=1.0\linewidth]{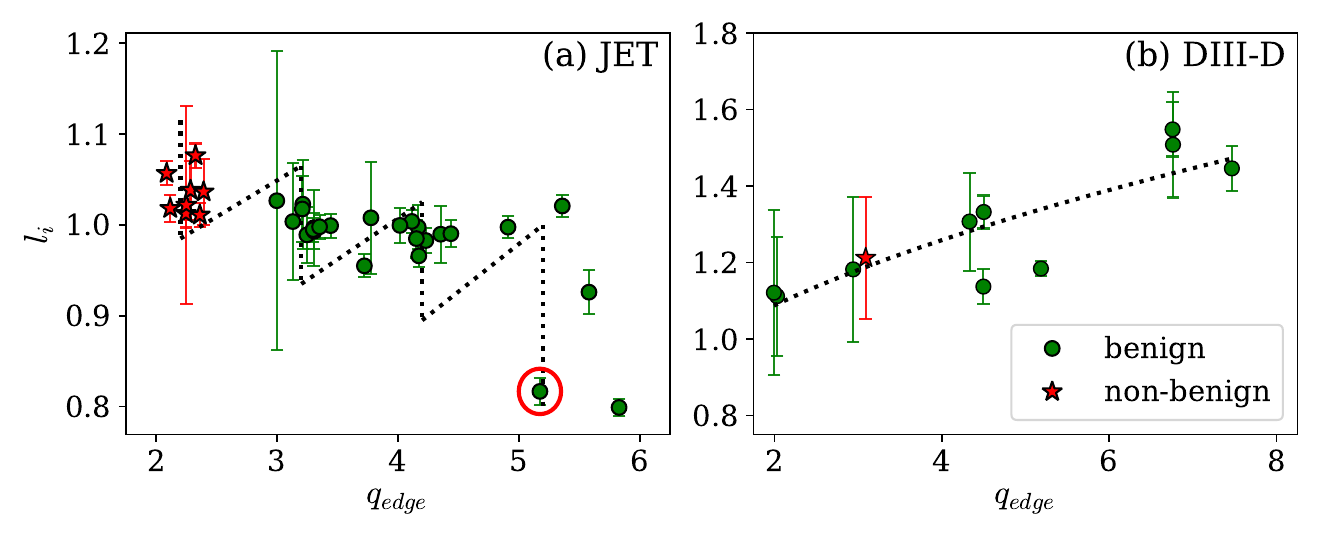}
    \caption{Internal inductance $l_i$ versus the edge safety factor $q_\text{edge}$ immediately before the terminating MHD event on JET (Panel a) and DIII-D (Panel b). Error bars reflect the systematic uncertainty on the $l_i$ measurement, mainly depending on the elongation $\kappa$, as discussed in the corresponding Section \ref{sec:phenomenology_termination} and Appendix \ref{sec:appendix_filtering_DIII_datapoints}. The black dotted lines recreate the empirical MHD stability boundary reported by Wesson \textit{et al.} \cite{Wesson_1989}. The encircled data point in Panel (a) corresponds to discharge \#95135, which was previously used as a reference case for benign termination at JET and is not found to be representative of the broader dataset.}
    \label{fig:li_vs_qedge_both_machines}
\end{figure}
In the previous section, it was suggested that RE current peaking could have a significant effect on MHD stability and the corresponding stability boundaries in $q_\text{edge} \approx 3$. To test this hypothesis, Fig.~\ref{fig:li_vs_qedge_both_machines} shows the internal inductance $l_i$ on the y-axis as a function of the edge safety factor $q_\text{edge}$ on the x-axis for JET (Panel a) and DIII-D (Panel b) for all studied discharges prior to the terminating MHD event. For JET, error bars represent the relative uncertainty in separating $l_i$ and $\beta_p$, which depends on the elongation $\kappa$, as discussed by Lao \textit{et al.} \cite{Lao_1985}. For DIII-D, error bars reflect the deviation of the measured values from analytically estimated values, with datapoints carrying uncertainties larger than $25\%$ excluded from this particular plot. Further details on this filtering procedure and the unfiltered $l_i$ values for DIII-D are provided in Appendix~\ref{sec:appendix_filtering_DIII_datapoints}.\par 
The variation in $l_i$ prior to termination is proposed to explain the observed differences in the terminating edge safety factors and could influence their characteristics and underlying physical mechanisms of the MHD events. The dashed lines in Fig.~\ref{fig:li_vs_qedge_both_machines} approximate different branches of the empirical kink stability diagram for rotating MHD modes in JET, as shown by Wesson \textit{et al.} \cite[Fig.\,6]{Wesson_1989}, drawing inspiration from the theoretical modeling results from Cheng \textit{et al.} \cite[Fig.\,4]{Cheng_1987}. The sawtooth-like behavior visible in Panel (a) corresponds to the external kink and tearing mode boundary in that paper. In contrast, the stability boundary identified for DIII-D aligns with the branch denoted as the “density limit disruption” by Wesson, which Cheng attributed to resistive internal kinks.\par

The hypothesis of stronger resistive effects on DIII-D is supported by the observation that nearly all benign cases exhibit small spikes in plasma current prior to the terminating MHD event. These spikes could be indicative of nonlinear coupling between ideal and resistive sideband modes, as discussed by Nardon \textit{et al.} \cite{Nardon_2023_PoP,Nardon_2023_NF}. While 14 out of 17 terminations on DIII-D exhibit such current spikes, only 8 out of 37 do so on JET. On DIII-D, 3 out of 4 cases without spikes are to non-benign terminations, whereas on JET no clear ordering parameter was identified that distinguishes cases with current spikes from those without.\par 

The red circle in Fig. \ref{fig:li_vs_qedge_both_machines}(a) markes the JET discharge \#95135. It was previously used for modeling purposes \cite{bandaru2021magnetohydrodynamic, reux2021prl}, and features a very small $l_i = 0.81$ (terminating at $q_{\text{edge}} \approx 5$). This is the second-lowest value observed and far from the majority of discharges (mean $0.99$ with a standard deviation of $0.05$). This is, on one hand, consistent with a broad current profile capable of producing a double tearing mode\cite{bandaru2021magnetohydrodynamic}, which relies on a RE current distribution peaking at mid-radius rather than on-axis. On the other hand, it shows that discharge \#95135 is nearly singular within the dataset, and therefore not representative for the termination dynamics on JET.\par 
Placing the measurements in this work into the broader context of previous studies suggests that JET and DIII-D likely encounter different stability boundaries due to differences in the precise formation of the RE current profile during the current quench. It is likely that machines such as DIII-D and ASDEX Upgrade behave more similarly to each other than to JET, since they show only limited avalanche runaway electron generation, while TCV shows almost none. Although the internal inductance has a strong influence on the termination parameters, it remains unclear from the current experimental data how the state of recombination affects or varies the current peaking, and whether this could link the recombination state to the nature of the terminating MHD event. It can be speculated that, for the high-$I_p$ cases on JET, the stronger current peaking together with the lower cross-sections creates a situation in which recombination is unsuccessful due to re-ionization. Such re-ionization has been described as possible for high RE current densities \cite{Hollmann2023,Hoppe2025} and has previously been observed on DIII-D in a similar scenario \cite{pazsoldan2021nf}.\par
\begin{figure}
    \centering
    \includegraphics[width=0.6\linewidth, trim=0 0 0 0, clip]{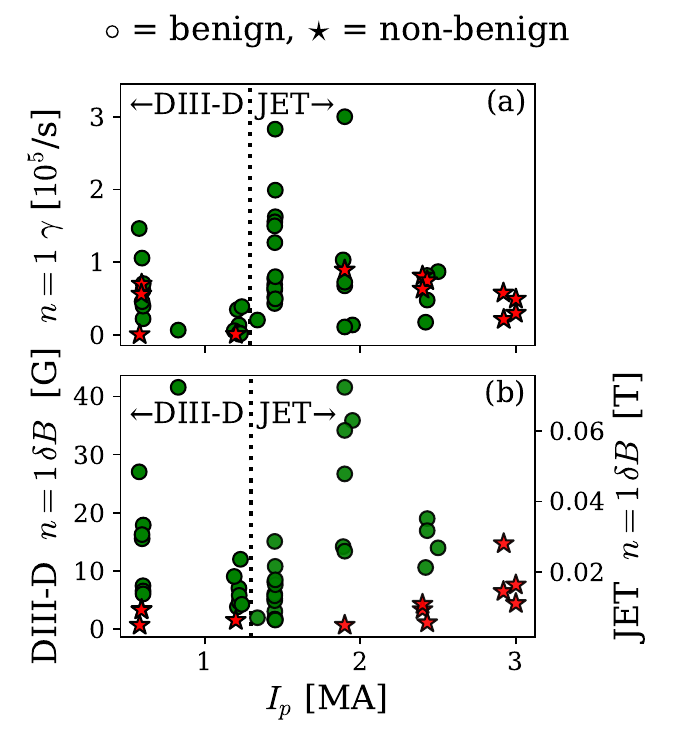}
    \caption{Measured growth rates (upper row) and perturbed $\delta B$ (lower row) of the terminating MHD events, for DIII-D (l.h.s.) and JET (r.h.s.) shown over the pre-disruptive plasma current. Growth rates do not allow for a separation of benign and non-benign cases while nearly all non-benign cases have lowest $\delta B$ values compared to their benign counterparts at similar $I_{p}$.}
    \label{fig:Ip_vs_gamma_dotB_vs_deltaB}
\end{figure}
So far, in this work, the amplitude of the terminating MHD events, predominantly the $n=1$ component of $\delta B$, has been used to discriminate between benign and non-benign cases. However, the exact mechanism linking the state of recombination to a benign or non-benign termination remains unclear so far. One hypothesis relates the free electron density $n_e$ to the Alfvénic timescale $\tau_A$, the timescale on which ideal MHD modes grow \cite{pazsoldan2019ppcf, sheikh2024ppcf}. Since benign cases indeed have a significantly lower $n_e$ than their non-benign, (re)ionized counterparts, their $\tau_A$ is shorter, allowing flux surfaces to open up more faster and, thus, more efficiently during the disruption \cite{Boozer_2016}. To test this hypothesis, the $n=1$ growth rates are sampled for both machines and are shown in Panel (a) of Fig.~\ref{fig:Ip_vs_gamma_dotB_vs_deltaB} as a function of the pre-disruptive plasma current.\par 
\begin{table}
\centering
\renewcommand{\arraystretch}{1.3} 
\begin{tabular}{r|c|c}
                        & \textcolor{OliveGreen}{\textbf{benign} }             & \textcolor{red}{\textbf{non-benign} }                 \\ \hline
hypothetical $n_e$ {[}1/m${}^{-3}{]}$   & $< 1 \cdot 10^{18}$ & $\approx 1 \cdot 10^{20}$ \\
hypothetical  $\tau_A$ {[}\textmu s{]}      & $< 0.03$            & $\approx 0.3$             \\
measured $\gamma^{-1}$ {[}\textmu s{]} & 11                  & 19                        \\
combined $\tau_A \cdot \gamma$   & $\approx 0.003$     & $\approx 0.015$ \\
\end{tabular}
    \caption{Hypothetical densities and Alfvén times for a recombined/benign and a non-recombined/non-benign termination scenario in JET. The inverse growth rate (typical time scale) $\gamma^{-1}$ corresponds to the average measured growth rates as shown in Fig. \ref{fig:Ip_vs_gamma_dotB_vs_deltaB}. Their product $\tau_A \cdot \gamma$ indicates a mixture of ideal and resistive effects.}
    \label{tab:growth_rate_comparison}
\end{table}
As shown, the exponential growth rates do not serve as a reliable discriminator between benign and non-benign cases on either machine, with both showing very similar absolute values on the order of micro-seconds. For both machines (DIII-D on the l.h.s., JET on the r.h.s., separated by vertical dotted lines), there is the trend of smaller growth rates for higher $I_p$. This suggests an inherent relation of the absolute value of $I_p$ before the primary disruption, the formation of the RE current profile immediately after the disruption, and its effect on the terminating MHD instability. Focusing on JET, one can calculate the Alfvénic timescale based on assumed electron densities, as shown in Tab.~\ref{tab:growth_rate_comparison}. A difference of two orders of magnitude in $n_e$ translates into roughly one order of magnitude difference in $\tau_A \sim \sqrt{n_e}$. However, the average inverse growth rates on JET vary only by a factor of two between benign and non-benign cases. This demonstrates that the hypothesis that recombination controls MHD growth purely through the Alfvén time is not supported by the data. Instead, the product $\tau_A \cdot \gamma$ suggests a resistive timescale, with values consistent with tearing modes or resistive kink modes, as discussed in more detail by Paz-Soldan \textit{et al.} \cite{pazsoldan2019ppcf}. The resistive time scales are not straightforward to calculate due to the absence of precise temperature measurements, but they are usually orders of magnitude larger than the Alfvén time.\par 
Panel (b) of Fig.~\ref{fig:Ip_vs_gamma_dotB_vs_deltaB} shows the corresponding $\delta B$ values. Most non-benign cases exhibit lower $\delta B$ compared to their benign counterparts at similar $I_{p}$. Measurements of the peak $\dot{B}$ show a similar trend to the integrated $\delta B$. For all cases studied, the $n=2$ components, where available, yield results consistent with the $n=1$.\par 
A possible theoretical interpretation linking the dominance of $\delta B$ to the benign character of terminations is that larger $\delta B$ provides a stronger seed for stochastization and island growth and overlap in the presence of resistive effects. One hypothesis is that external kink modes alone may only open a localized region of the field structure, leading to focused RE losses and measurable localized wall impacts \cite{Boozer_2016}. In contrast, nonlinear interactions of (resistive) external kink and tearing modes can generate widespread stochasticity of the magnetic field, allowing uniform RE deconfinement. A full interpretation of these observations requires nonlinear and resistive MHD modeling.\par

\section{MHD modeling of the termination boundaries}
\label{sec:modeling}
For the purpose of interpreting the MHD boundaries and their characteristics observed in this work, dedicated linear and resistive MHD calculations were carried out with the CASTOR3D code \cite{Strumberger_2017,Strumberger_2019}.\par 
As an experimental starting point, two representative, recombined, benign discharges were chosen from the experimental data set just prior to termination, at a time when all relevant quantities and the plasma equilibrium were still stable. For JET, discharge \#95170 at $48.67$\,s ($B_t = 3$\,T, $q_\text{edge} = 3.2$, $l_i = 0.98$, $I_{RE} = 1.27$\,MA) was selected, and for DIII-D, discharge \#184604 at $0.867$\,s ($B_t = 2.2$\,T, $q_\text{edge} = 3.2$, $l_i = 1.142$, $I_{RE} = 0.475$\,MA) was used. For a recombined, post-disruptive companion plasma, the density is below $1 \cdot 10^{18}~\mathrm{m^{-3}}$ and temperatures are in the single-digit eV range \cite{pazsoldan2019ppcf}. Dedicated measurements of current and loop voltage in both reference discharges suggest resistivity values of $\eta \approx 5 \cdot 10^{-5}~\Omega\,$m for both machines. This value is in agreement with previous measurements for such plasma conditions \cite{pazsoldan2019ppcf}, and its rather high value results from the very low temperatures after the thermal quench, calling into question the basic MHD assumption of complete ionization.\par
\begin{figure}
    \centering
    \includegraphics[width=0.5\linewidth]{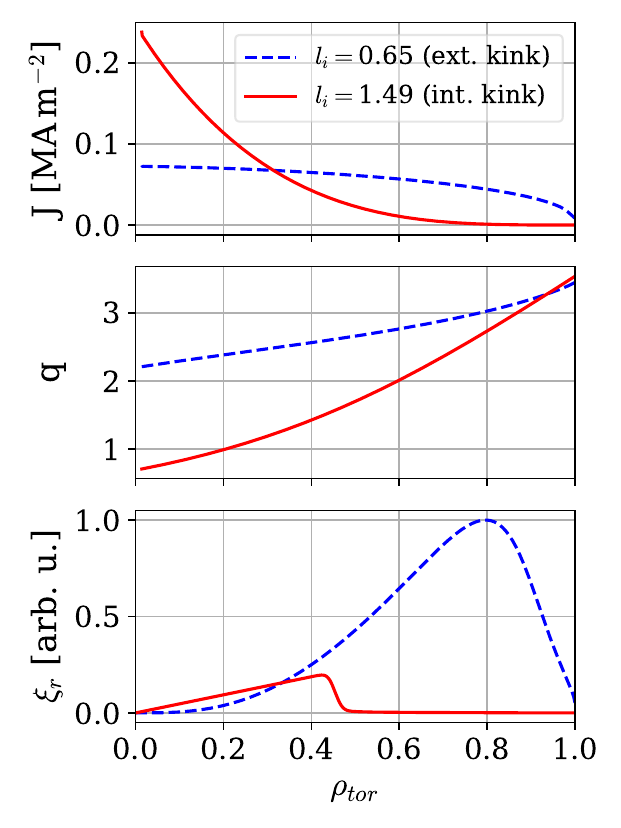}
    \caption{Examples from the CASTOR3D modeling, varying the current density (Panel~a) for fixed edge values of the safety-factor profile (Panel~b) to scan different $l_i$ values. The $l_i$ values for which the curves are shown are just beyond the upper, respectively lower stability boundary. As can be seen in Panel~(c), the mode eigenfunctions suggest different instability natures, with strong, external kink modes ($m/n = 4/1$) for low $l_i$ values (blue dashed line) and weaker, internal kink modes ($m/n = 1/1$) for high $l_i$ values (red solid line). A resistivity of $10^{-5}~\Omega\,$m is used.}
    \label{fig:jqeigefunction_vs_radius}
\end{figure}

 
As the internal inductance and edge safety factor were found in the experiment to determine the MHD stability boundaries, both were varied in the following scans to track their effect. To generate otherwise self-similar equilibria, the NEMEC equilibrium solver was used \cite{HIRSHMAN_NEMEC}. As an example, the resulting current and safety-factor profiles are shown in Fig.~\ref{fig:jqeigefunction_vs_radius}(a) and (b) as a function of the square root of the normalized toroidal flux $\rho_\text{tor}$. The depicted $l_i$ values are representative of cases just beyond the upper and lower stability boundaries found for $q_\text{edge}=3.2$, with the flat current profile for low $l_i=0.65$ (dashed blue line) and the more peaked one for $l_i=1.49$ (solid red line). The corresponding instability eigenfunctions from the MHD modeling in Panel (c) suggest strong, external kink modes (peaking outside mid-radius) for low $l_i$ values and weaker, internal kink modes (peaking inside mid-radius) for higher $l_i$ values. The internal kink is numerically identified as an $m/n = 1/1$ mode, while the external kink is identified as an $m/n = 4/1$ mode.\par
Finding internal kink modes seems to be a trivial result, given that the core safety factors drop below $1$, see Fig. \ref{fig:jqeigefunction_vs_radius}(b). Therefore, to improve the understanding of the upper stability boundary, another scan of current profiles was performed in which the safety factor was enforced to remain larger than $1$ for high $l_i$ values. In this scan, the internal kink disappeared and only a small, underlying, external resistive mode was found, without a clear instability boundary. Therefore, in the linear, resistive picture, the existence of an upper boundary, as observed in the experiments, relies necessarily on the assumption that the core safety factors can fall below $1$. Owing to the absence of neutral beam injection and the Motional Stark Effect (MSE) diagnostic, this modeling result cannot currently be validated experimentally. However, it represents the most plausible scenario, as the experimentally observed upper kink stability cannot otherwise be reproduced by the MHD modeling. Internal kink modes are assumed to serve as a seed for (resistive) non-linear events, resulting in stochastization and, hence, deconfinement of the RE beam. Modeling such mode interactions would require nonlinear MHD calculations, which are the subject of future work and cannot be applied for scanning large parameter spaces due to their computational expense.\par
\begin{figure}
    \centering
    \includegraphics[width=0.66\linewidth]{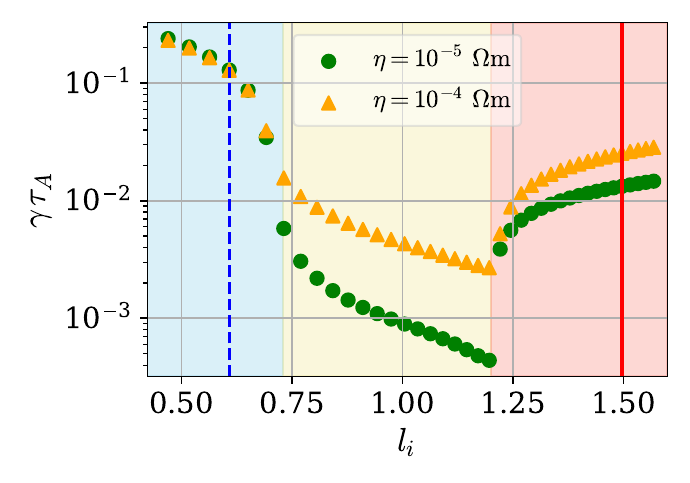}
    \caption{Normalized growth rates as a function of the tested $l_i$ values for a fixed $q_{\text{edge}} = 3.2$ as calculated by the CASTOR3D code. One can see the rise of the instability for the lowest and highest $l_i$ values and a rather stable region in between (yellow), with growth rates below $\gamma\,\tau_A < 10^{-2}$. Results for two different resistivities are shown. The instability boundary at high $l_i$ values, identified as being governed by internal kink modes (red region), is enhanced by resistive effects but remains less rapid than the external kink modes found at the low-$l_i$ boundary (blue region), which are nearly independent of resistive effects.}
    \label{fig:modeling_gamma_vs_li}
\end{figure}

In addition, the effect of using different resistivities in the MHD calculations was investigated. For this purpose, the linear, normalized growth rates $\gamma\,\tau_A$ were scanned for the DIII-D reference case over a range of $l_i$ values, as shown in Fig.~\ref{fig:modeling_gamma_vs_li}. One can see that using different resistivities modifies the calculated growth rates only for the high $l_i$ values identified with internal kink modes (red region). However, no new types of instabilities nor shifts of the stability boundaries are observed. In this linear picture, the external–kink events (blue region) are significantly more rapid in nature, yielding larger growth rates. A rather stable region is found for medium $l_i$ values (see yellow region), where normalized growth rates are below 1\%, and in which stabilizing effects, such as flows or finite Larmor radius effects, can suppress such slowly growing modes. While this analysis is based on the growth rates, a realistic calculation of $\delta B$ values would require more sophisticated, non-linear MHD modeling with complex synthetic diagnostics and is the subject of future work. In addition, such future work could explore whether flatter current profiles and, consequently, flatter $q$-profiles could lead to more global mode structure, beneficial for efficient stochastication. A direct comparison with the growth rates assessed experimentally, e.g., in Tab.~\ref{tab:growth_rate_comparison} or Fig.~\ref{fig:Ip_vs_gamma_dotB_vs_deltaB}, shows that the predicted mode growth is faster than what is observed experimentally. This suggests that there are effects damping the mode growth in reality, such as nonlinear resistive effects or stabilizing influences from the wall, which are not included in the current modeling. The vertical lines in Fig. \ref{fig:modeling_gamma_vs_li} correspond to the $l_i$ values used in Fig.~\ref{fig:jqeigefunction_vs_radius} and shown in the corresponding linestyle and color.\par 
\begin{figure}
    \centering
    \includegraphics[width=0.85\linewidth]{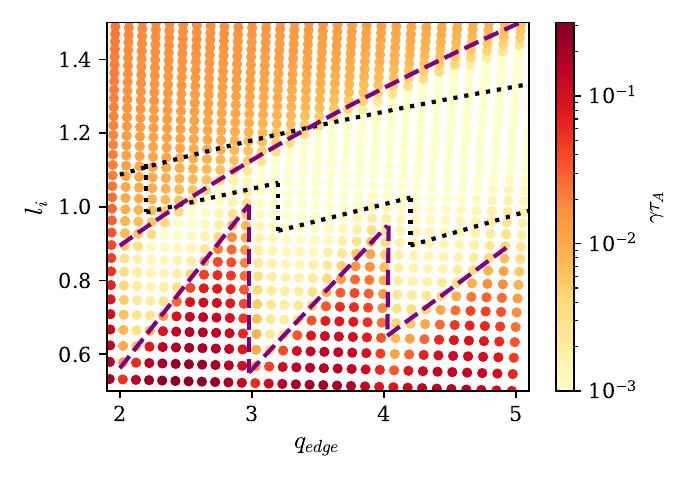}
    \caption{MHD modeling using the linear, resistive CASTOR3D code shows the normalized instability growth rates (color-coded) in the kink stability space with the edge safety factor on the x-axis and the internal inductance on the y-axis. The parameter scan is based on DIII-D discharge \#184604. A purple dashed line is included to enhance the visibility of the stable (bright yellow) and unstable (dark orange and red) regions. The dotted line corresponds to the experimental stability boundary shown in Fig.~\ref{fig:li_vs_qedge_both_machines}. A resistivity of $10^{-5}~\Omega\,$m is used.}
    \label{fig:modeling_parameter_space}
\end{figure}
Figure~\ref{fig:modeling_parameter_space} shows the full scan in kink stability space. This was plotted for the DIII-D case, as corresponding JET simulations in all presented aspect produced nearly identical results. This agreement between the two machines is expected since kink stability relies on normalized, dimensionless quantities and the precise equilibrium shape or absolute current values play a minor role. As shown in the figure, all relevant instability boundaries observed in the experiment (dotted lines) are qualitatively reproduced in the modeling (dashed line). While the simulated sawteeth structures are generally larger than observed experimentally, this can be explained by the realistic trajectory of the experiment, which typically hits the uppermost part of each sawtooth when lowering the edge safety factor. In addition, the exact slope of the upper termination boundary strongly depends on the exact shape of the assumed current profile, which cannot be measured experimentally. An exact match between experiment and theory is not expected, as the equilibria generated from a single reference case cannot perfectly represent the experimental ensemble of discharges.\par 

Overall, this modeling effort demonstrates that the observed termination boundaries are fully consistent with MHD theory and the presence of kink modes. Resistive effects modify the growth rates of the upper stability boundary, but do not significantly change the boundary structures. Internal kinks are found dominant for the termination boundary associated with the DIII-D datapoints at high $l_i$, whereas external kinks play a larger role for JET at lower $l_i$, at least within this linear MHD picture. Nonlinear, resistive, extended-MHD calculations are not feasible for parameter scans and, in future work, can instead focus on studying individual points in the parameter space.

\section{Summary and Outlook}
\label{sec:summary_outlook}

This study presents a comparative analysis of benign and non-benign termination of post-disruptive RE beams in JET and DIII-D following hydrogenic secondary injections. It relies on fast magnetic measurements and equilibrium reconstructions to characterize the evolution and termination of the RE beam.\par 
In the JET dataset, non-benign terminations at high pre-disruptive plasma current occur at low edge safety factor values around $q_{\text{edge}} = 2$ and are preceded by intermittent, non-terminating MHD activity at higher rational $q_{\text{edge}}$. These cases are characterized by short RE plateau phases, smaller plasma cross-sections, higher RE current densities, and more peaked current profiles (reflected in larger internal inductance $l_i$) prior to the termination. These factors may prevent recombination in the high-$I_p$ JET cases due to re-ionization of the background plasma, leading to a larger number of non-benign terminations at these $I_p$ values. Benign terminations on JET typically occur at higher edge safety factor, $q_{\text{edge}} \geq 3$. The DIII-D dataset spans a broader range of terminating edge safety factors, with VDE terminations occurring at higher $q_\text{edge}$. Benign center-post terminations are predominantly found at the lowest $q_{\text{edge}} \approx 2$, while non-benign cases show a tendency toward higher $q_{\text{edge}}$.\par
Across both machines, the RE current peaking emerges as the dominant factor in determining at which edge safety factors the termination occurs. This is compatible with the concept of the kink stability space and both machines are found to terminate at different stability boundaries. Dedicated modeling effort with the linear, resistive CASTOR3D MHD code is able to explain the measured termination boundaries with the presence of resistive internal (DIII-D) and external (JET) kink modes and bolsters the presented interpretation of the experimental phenomenology. It is likely that the precise mechanisms of RE generation produce systematically different RE current distributions in both machines~\cite{Smith_2010,Loarte_2011}. This, in turn, then leads to the observed termination in different MHD stability boundaries. Finally, it was shown that the JET discharge \#95135 is of a singular nature in terms of edge safety factor and current peaking, and not representative of the majority of observed JET terminations in this work. \par 
Measured MHD growth rates of the terminating $n=1$ activity do not sufficiently distinguish between benign and non-benign cases, indicating that the Alfvén time, on which ideal MHD grows and which is linked to the free-electron density, cannot separate recombined, benign terminations from ionized, non-benign ones. Instead, the decisive ordering parameter across both machines is the amplitude of the magnetic perturbation at the moment of loss: non-benign terminations consistently exhibit low $\delta B$. The precise understanding of this observation is likely requires the study of nonlinear coupling between ideal and resistive dynamics which govern whether RE losses are spatially distributed or localized.\par 
Several open questions and directions for future research emerge from this analysis. First, classification schemes for benign and non-benign terminations differ across devices, currently relying on various indicators such as IR camera data, neutron or hard X-ray spikes, and the current quench rate. A standardized approach would facilitate more consistent cross-machine comparisons.\par 

Although the agreement between experimentally observed termination boundaries and MHD modeling increases confidence in the models currently in use, it also supports that detailed modeling of RE current–profile formation must be part of the self-consistent extrapolation of MHD behavior during benign terminations in future reactor scenarios. The difficulties encountered in the high-current JET cases could be addressed by actively steering the current-profile evolution in the RE plateau phase toward values that allow termination at higher edge safety factors, thereby avoiding compression to low cross-sections and the subsequent re-ionization.\par 

Finally, the underlying roles of recombination and re-ionization in determining the benign or non-benign character of a termination remain unresolved. More advanced MHD modeling of unsuccessful, non-benign terminations will be essential for clarifying this connection, emphasizing the need for nonlinear resistive MHD simulations that self-consistently incorporate RE profile formation, recombination dynamics, and termination processes to enable reliable extrapolation. In particular, the author encourages broader exploitation of the available JET dataset.

\appendix
\section{Appendix}
\subsection{Filtering of DIII-D data points based on $l_i$ values}
\label{sec:appendix_filtering_DIII_datapoints}
\begin{figure}[H]
    \centering
    \includegraphics[width=0.66\linewidth]{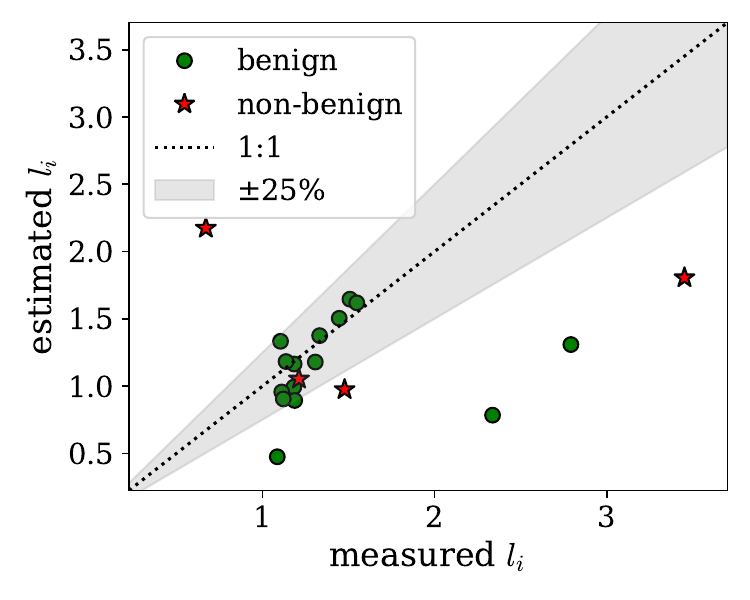}
    \caption{For the studied DIII-D dataset, the measured $l_i$ values are compared with the estimated $l_i$ values obtained using the approach described in Appendix~\ref{sec:appendix_li_estimates}. A subset of datapoints showed deviations greater than $25\%$, which are excluded from the $l_i$ values plotted for DIII-D in Fig.~\ref{fig:li_vs_qedge_both_machines} as their values are likely to be unphysical.}
    \label{fig:li_estimate_vs_EFIT}
\end{figure}
As shown by Lao \textit{et al.} \cite[Fig.~2]{Lao_1985}, $l_i$ and $\beta_p$ cannot be reliably separated for circular equilibria with elongations of $\kappa < 1.1$. This condition applies to nearly all center-post compression terminations on DIII-D, but not to cases with VDEs ($\kappa=1.3-1.7$) or to most JET discharges, which typically have $\kappa > 1.1$.\par
The predicted relative deviation between the "true" and the "measured" $l_i$ values as a function of $\kappa$, as reported by Lao \textit{et al.}, is used to inform the error bars on the JET $l_i$ measurements in Fig.~\ref{fig:li_vs_qedge_both_machines}. However, some data points in the DIII-D dataset exhibit unphysically high or low $l_i$ values, which complicates reconstruction of the stability boundary.\par
Accordingly, while Fig.~\ref{fig:IRE_vs_qedge_both_machines}(b) shows 17 datapoints for $I_{RE}$ versus $q_\text{edge}$, six points were removed in the $l_i$ versus $q_\text{edge}$ plot in Fig.~\ref{fig:li_vs_qedge_both_machines}. The excluded points are those for which the estimated $l_i$ (see Appendix~\ref{sec:appendix_li_estimates}) deviated by more than $25\%$ from the measured values. Notably, this affects three out of four non-benign cases, which is consistent with the expectation that the isotropic pressure assumption in the derivation of the estimated $l_i$ values likely breaks down in non-benign cases.
\subsection{Estimating $l_i$ for cases with low $\kappa$}
\label{sec:appendix_li_estimates}

Looking at Eq. (16) from Cooper \textit{et al.} \cite{Cooper_1982}, derived from Shafranov \textit{et al.} \cite{Shafranov_1971},
\begin{equation*}
    \frac{1}{2}(\beta_{p\perp}+\beta_{p\parallel})+W_{p\phi}+\frac{1}{2}W_{p\perp}+\frac{l_i}{2}=\frac{S_1}{2}+S_2 \text{,}
\end{equation*}
with 
\begin{itemize}
    \item the perpendicular and parallel component of $\beta_p$, $\beta_{p\perp}$ and $\beta_{p\parallel}$, which can be summarized in the case of isotropy into $\frac{1}{2}(\beta_{p\perp}+\beta_{p\parallel})=\beta_p$,
    \item the flow energy correction terms $W_{p\phi}$ and $W_{p\perp}$, which, typically, give corrections of 10\% and, in absence of external torque and for the low ion temperatures in our case, are set to zero,
    \item the geometry integrals $S_1$ and $S_2$.
\end{itemize}
The geometry integrals are constant for a certain time point and not affected by the degeneracy when solving for $l_i$ and $\beta_p$, see Lao \textit{et al.} \cite{Lao_1985}. Under this condition, the measured, degenerate $\beta_p+\frac{l_i}{2}=const.$ from cases with $\kappa < 1.1$ can still be considered correct. In principle, 
\begin{equation*}
    \beta_p = \frac{2 \mu_0 \overline{p}}{\overline{B_p^2}} \text{,}
\end{equation*}
with the average
\begin{equation*}
    B_p = \frac{\mu_0 I_{RE}}{l}
\end{equation*}
with, for a given aspect ratio $\kappa$ and cross section $A$,
\begin{equation*}
    l \approx \sqrt{\pi A/\kappa}(3(\kappa+1)-\sqrt{(3\kappa+1)(\kappa+3)}) \text{.}
\end{equation*}
Due to their large energies, the average pressure $\overline{p}$ is dominated by the REs in case of a purged background plasma, such that for an isotropic pressure distribution
\begin{equation*}
    \overline{p} = \frac{I_{RE}}{eA}\cdot m_e c \gamma \beta \text{.}
\end{equation*}
This ansatz is similiar to what has been done in the work of Loarte \textit{et al.} \cite{Loarte_2011}, based on Refs. \cite{Knoepfel_1979,Fujita_1991,Kuznetsov_2004}. Here, the relativistic corrections are the Lorentz factor
\begin{equation*}
   \gamma = 1+ \frac{E_{RE}}{m_e c^2}  
\end{equation*}
with $m_e c^2 \approx 0.511$\,MeV and $E_{RE}$ the kinetic energy of the REs. $\beta$ defines the ratio of the RE velocity to the speed of light,
\begin{equation*}
    \beta = \frac{v}{c} = \sqrt{1-\frac{1}{\gamma^2}} \text{.}
\end{equation*}
For an average population of 3.5 MeV REs, $\gamma \approx 7.85$ and $\beta \approx 0.99$ and, therefore, 
\begin{equation*}
    \overline{p} \approx 7.8\cdot\frac{I_{RE}}{eA}\cdot m_e c\text{.}
\end{equation*}
A fully "predictive" $\beta_p$ is given by
\begin{equation*}
    \beta_p \approx 15.6 \cdot \frac{m_e\,c\,l(\kappa,A)^2}{e\,A\,\mu_0\,I_{RE}} \text{.}
\end{equation*}
The average population energy is chosen such that the measured and estimated values match for the VDE cases with high $\kappa=1.3-1.7$ in which the separation of $\beta_p$ and $l_i$ is feasible. They agree within $10$ \%, which is the accuracy expected for the corresponding $\kappa$ values. While the energy therefore serves as a linear fudge factor, the used value is still very physical and can be seen as an indirect measurement of the RE energy. It is expected that non-purged cases do not allow to assume an isotropic pressure distribution, such that this analysis is likely to break down.\par

Following the approach to measure $\beta_{p,meas.}+l_{i,meas.}/2=S_1/2+S_2$ and estimate $l_{i,est.}$ based on an estimated $\beta_{p,est.}$, the internal inductance can be written as 
\begin{equation*}
    l_{i,est.} = 2\cdot (S_1/2+S_2-15.6 \cdot \frac{m_e\,c\,l(\kappa)^2}{e\,A\,\mu_0\,I_{RE}}) \text{.}
\end{equation*}

\printbibliography

@article{pazsoldan2019ppcf,
  author       = {C. Paz-Soldan and others},
  title        = {},
  journal      = {Plasma Physics and Controlled Fusion},
  volume       = {61},
  pages        = {054001},
  year         = {2019}
}

@article{Nardon_2023_NF,
doi = {10.1088/1741-4326/acc417},
url = {https://doi.org/10.1088/1741-4326/acc417},
year = {2023},
publisher = {IOP Publishing},
volume = {63},
number = {5},
pages = {056011},
author = {Nardon, E. and Särkimäki, K. and Artola, F.J. and Sadouni, S. and team, the JOREK and Contributors, JET},
title = {On the origin of the plasma current spike during a tokamak disruption and its relation with magnetic stochasticity},
journal = {Nuclear Fusion}
}

@article{Nardon_2023_PoP,
    author = {Nardon, E. and Bandaru, V. and Hoelzl, M. and Artola, F. J. and Maget, P. and JOREK Team and JET Contributors},
    title = {Non-linear dynamics of the double tearing mode},
    journal = {Physics of Plasmas},
    volume = {30},
    number = {9},
    pages = {092502},
    year = {2023},
    issn = {1070-664X},
    doi = {10.1063/5.0162608},
    url = {https://doi.org/10.1063/5.0162608},
    eprint = {https://pubs.aip.org/aip/pop/article-pdf/doi/10.1063/5.0162608/18107820/092502\_1\_5.0162608.pdf},
}

@inproceedings{GradRubin1958,
  author       = {Grad, H. and Rubin, H.},
  title        = {Hydromagnetic Equilibria and Force-Free Fields},
  booktitle    = {Proceedings of the Second United Nations International Conference on the Peaceful Uses of Atomic Energy},
  year         = {1958},
  volume       = {31},
  pages        = {190--197},
  publisher    = {International Atomic Energy Agency},
  address      = {Geneva},
}

@article{Shafranov1966,
  author       = {Shafranov, V. D.},
  title        = {Plasma Equilibrium in a Magnetic Field},
  journal    = {Reviews of Plasma Physics},
  volume       = {2},
  pages        = {103--151},
  year         = {1966},
  publisher    = {Consultants Bureau},
  address      = {New York},
}

@article{Cooper_1982,
doi = {10.1088/0032-1028/24/9/014},
url = {https://doi.org/10.1088/0032-1028/24/9/014},
year = {1982},
publisher = {},
volume = {24},
number = {9},
pages = {1183},
author = {W. A. Cooper and A. J. Wootton},
title = {βp analysis for Tokamak plasma with anisotropic pressure and mass flow},
journal = {Plasma Physics}
}

@article{Shafranov_1971,
doi = {10.1088/0032-1028/13/9/006},
url = {https://doi.org/10.1088/0032-1028/13/9/006},
year = {1971},
publisher = {},
volume = {13},
number = {9},
pages = {757},
author = {V. D. Shafranov},
title = {Determination of the parameters βI and li in a Tokamak for arbitrary shape of plasma pinch cross-section},
journal = {Plasma Physics}
}

@article{Cheng_1987,
doi = {10.1088/0741-3335/29/3/006},
url = {https://dx.doi.org/10.1088/0741-3335/29/3/006},
year = {1987},
publisher = {},
volume = {29},
number = {3},
pages = {351},
author = {C. Z. Cheng and H P Furth and A H Boozer},
title = {MHD stable regime of the Tokamak},
journal = {Plasma Physics and Controlled Fusion}
}

@article{Wesson_1989,
doi = {10.1088/0029-5515/29/4/009},
url = {https://dx.doi.org/10.1088/0029-5515/29/4/009},
year = {1989},
publisher = {},
volume = {29},
number = {4},
pages = {641},
author = {Wesson, J. A. and Gill, R.D. and Hugon, M. and Schüller, F.C. and Snipes, J.A. and Ward, D.J. and Bartlett, D.V. and Campbell, D.J. and Duperrex, P.A. and Edwards, A.W. and Granetz, R.S. and Gottardi, N.A.O. and Hender, T.C. and Lazzaro, E. and Lomas, P.J. and Lopes Cardozo, N. and Mast, K.F. and Nave, M.F.F. and Salmon, N.A. and Smeulders, P. and Thomas, P.R. and Tubbing, B.J.D. and Turner, M.F. and Weller, A.},
title = {Disruptions in JET},
journal = {Nuclear Fusion}
}

@article{Rosenbluth_1997,
doi = {10.1088/0029-5515/37/10/I03},
url = {https://doi.org/10.1088/0029-5515/37/10/I03},
year = {1997},
publisher = {},
volume = {37},
number = {10},
pages = {1355},
author = {M. N. Rosenbluth and S. V. Putvinski},
title = {Theory for avalanche of runaway electrons in tokamaks},
journal = {Nuclear Fusion}
}

@article{Munaretto_2021,
    author = {Munaretto, S. and Strait, E. J. and Logan, N. C.},
    title = {Optimizing the differential connection schemes for detecting 3D magnetic perturbations in DIII-D},
    journal = {Review of Scientific Instruments},
    volume = {92},
    number = {7},
    pages = {073504},
    year = {2021},
    issn = {0034-6748},
    doi = {10.1063/5.0045453},
    url = {https://doi.org/10.1063/5.0045453},
    eprint = {https://pubs.aip.org/aip/rsi/article-pdf/doi/10.1063/5.0045453/16017503/073504_1_online.pdf},
}

@article{Lao_1985,
doi = {10.1088/0029-5515/25/10/004},
url = {https://doi.org/10.1088/0029-5515/25/10/004},
year = {1985},
publisher = {},
volume = {25},
number = {10},
pages = {1421},
author = {Lao, L. L. and John, H. St. and Stambaugh, R.D. and Pfeiffer, W.},
title = {Separation of β̄p and ℓi in tokamaks of non-circular cross-section},
journal = {Nuclear Fusion}
}

@article{zohm2015magnetohydrodynamic,
  title={Magnetohydrodynamic stability of tokamaks},
  author={Zohm, H.},
  year={2015},
  journal={John Wiley \& Sons},
pages = {ISBN: 978-3527412327}
}

@article{igochine2015active,
  title={Active control of magneto-hydrodynamic instabilities in hot plasmas},
  author={Igochine, V. and others},
  year={2015},
  journal={Springer},
pages = {ISBN: 978-3662442210}
}

@article{freidberg1982ideal,
  title={Ideal magnetohydrodynamic theory of magnetic fusion systems},
  author={Freidberg, J. P.},
  journal={Reviews of Modern Physics},
  volume={54},
  number={3},
  pages={801},
  year={1982},
  publisher={APS}
}

@article{Chiu_1998,
doi = {10.1088/0029-5515/38/11/309},
url = {https://doi.org/10.1088/0029-5515/38/11/309},
year = {1998},
publisher = {},
volume = {38},
number = {11},
pages = {1711},
author = {S. C. Chiu and M.N. Rosenbluth and R.W. Harvey and V.S. Chan},
title = {Fokker-Planck simulations mylb
of knock-on electron runaway avalanche 
and bursts in tokamaks},
journal = {Nuclear Fusion}
}

@article{Dreicer_1959,
  title = {Electron and Ion Runaway in a Fully Ionized Gas. I},
  author = {Dreicer, H.},
  journal = {Physical Review},
  volume = {115},
  issue = {2},
  pages = {238--249},
  numpages = {0},
  year = {1959},
  publisher = {American Physical Society},
  doi = {10.1103/PhysRev.115.238},
  url = {https://link.aps.org/doi/10.1103/PhysRev.115.238}
}

@article{Breizman_2019,
doi = {10.1088/1741-4326/ab1822},
url = {https://doi.org/10.1088/1741-4326/ab1822},
year = {2019},
publisher = {IOP Publishing},
volume = {59},
number = {8},
pages = {083001},
author = {Breizman, B. N. and Aleynikov, Pavel and Hollmann, Eric M. and Lehnen, Michael},
title = {Physics of runaway electrons in tokamaks},
journal = {Nuclear Fusion}
}

@article{Loarte_2011,
doi = {10.1088/0029-5515/51/7/073004},
url = {https://doi.org/10.1088/0029-5515/51/7/073004},
year = {2011},
publisher = {},
volume = {51},
number = {7},
pages = {073004},
author = {Loarte, A. and Riccardo, V. and Martin-Solís, J.R. and Paley, J. and Huber, A. and Lehnen, M. and JET EFDA Contributors},
title = {Magnetic energy flows during the current quench and termination of disruptions with runaway current plateau formation in JET and implications for ITER},
journal = {Nuclear Fusion}
}

@article{liu2019nf,
  author       = {Y. Q. Liu and others},
  title        = {},
  journal      = {Nuclear Fusion},
  volume       = {59},
  pages        = {126021},
  year         = {2019}
}

@article{pazsoldan2021nf,
  author       = {C. Paz-Soldan and others},
  title        = {},
  journal      = {Nuclear Fusion},
  volume       = {61},
  pages        = {116058},
  year         = {2021}
}

@article{sheikh2024ppcf,
  author       = {U. Sheikh and others},
  title        = {},
  journal      = {Plasma Physics and Controlled Fusion},
  volume       = {66},
  pages        = {035003},
  year         = {2024}
}

@article{reux2021prl,
  author       = {C. Reux and others},
  title        = {},
  journal      = {Physical Review Letters},
  volume       = {126},
  pages        = {175001},
  year         = {2021}
}

@article{reux2022ppcf,
  author       = {C. Reux and others},
  title        = {},
  journal      = {Plasma Physics and Controlled Fusion},
  volume       = {64},
  pages        = {034002},
  year         = {2022}
}

@article{Smith_2010,
    author = {Smith, H. and Helander, P. and Eriksson, L.-G. and Anderson, D. and Lisak, M. and Andersson, F.},
    title = {Runaway electrons and the evolution of the plasma current in tokamak disruptions},
    journal = {Physics of Plasmas},
    volume = {13},
    number = {10},
    pages = {102502},
    year = {2006},
    issn = {1070-664X},
    doi = {10.1063/1.2358110},
    url = {https://doi.org/10.1063/1.2358110},
    eprint = {https://pubs.aip.org/aip/pop/article-pdf/doi/10.1063/1.2358110/15623648/102502\_1\_online.pdf},
}

@article{Hoppe2025,
  author = {M. Hoppe and J. Decker and U. Sheikh and S. Coda and C. Colandrea and B. Duval and O. Ficker and P. Halldestam and S. Jachmich and M. Lehnen},
  title = {An Upper Pressure Limit for Low-Z Benign Termination of Runaway Electron Beams in TCV},
  journal = {Plasma Physics and Controlled Fusion},
  volume = {67},
  pages = {045015},
  year = {2025}
}

@article{Hollmann2023,
  author = {E. M. Hollmann et al},
  title = {Trends in Runaway Electron Plateau Partial Recombination by Massive H2 or D2 Injection in DIII-D and JET and First Extrapolations to ITER and SPARC},
  journal = {Nuclear Fusion},
  volume = {63},
  pages = {036011},
  year = {2023}
}

@inbook{reux2024runaway,
title = "The Runaway Electron Benign Termination Scenario: Physics Processes and Operational Limits",
author = "C. Reux and U. Sheikh and C. Paz-Soldan and O. Ficker and M. Lehnen and S. Jachmich and S. Silburn and Lomas, \{P. J.\} and C. Lowry and N. Schoonheere and D. Craven and J. Wilson and M. Nocente and Molin, \{A. Dal\} and G. Szepesi and D. Kos and A. Boboc and A. Lvovskiy and M. Baruzzo and Antti Hakola and E. Joffrin and C. Sommariva and A. Battey and D. Brunetti and P. Buratti and H. Choudhury and J. Decker and N. Eidietis and M. Hoppe and H. Isliker and E. Kowalska-Strzeciwilk and G. Marcer and E. Nardon and V. Plyusnin and D. Rigamonti and L. Spolladore and E. Tomesova and M. Zerbini and \{JET Contributors\} and \{EUROfusion Tokamak Exploitation Team\}",
year = "2024",
publisher = "European Physical Society",
editor = "J. Kirk and L. Volpe",
booktitle = "50th EPS Conference on Plasma Physics",
url = "https://lac913.epfl.ch/epsppd3/2024/html/index.html"
}

@article{Knoepfel_1979,
doi = {10.1088/0029-5515/19/6/008},
url = {https://doi.org/10.1088/0029-5515/19/6/008},
year = {1979},
publisher = {},
volume = {19},
number = {6},
pages = {785},
author = {Knoepfel, H. and Spong, D. A.},
title = {Runaway electrons in toroidal discharges},
journal = {Nuclear Fusion}
}

@article{marini2024runaway,
  title={Runaway electron plateau current profile reconstruction from synchrotron imaging and Ar-II line polarization angle measurements in DIII-D},
  author={Marini, C. and Hollmann, Eric M and Tang, Shawn Wenjie and Herfindal, JL and Shiraki, Daisuke and Wilcox, Robert S and Del-Castillo-Negrete, Diego and Yang, Minglei and Eidietis, N and Hoppe, Mathias},
  journal={Nuclear Fusion},
  volume={64},
  number={7},
  pages={076039},
  year={2024},
  publisher={IOP Publishing}
}

@article{Fujita_1991,
author = {Fujita, T. and Fuke ,Yasutaka and Yoshida ,Zensho and Inoue ,Nobuyuki and Tanihara ,Takeo and Mori ,Ken-ichi and Fukao ,Masayuki and Tomita ,Yukihiro and Mohri ,Akihiro},
title = {High-Current Runaway Electron Beam in a Tokamak Plasma},
journal = {Journal of the Physical Society of Japan},
volume = {60},
number = {4},
pages = {1237-1246},
year = {1991},
doi = {10.1143/JPSJ.60.1237},
URL = { 
        https://doi.org/10.1143/JPSJ.60.1237
},
eprint = {    
        https://doi.org/10.1143/JPSJ.60.1237
}}

@article{Kuznetsov_2004,
doi = {10.1088/0029-5515/44/5/007},
url = {https://doi.org/10.1088/0029-5515/44/5/007},
year = {2004},
publisher = {},
volume = {44},
number = {5},
pages = {631},
author = {Y. K. Kuznetsov and R.M.O. Galvão and V. Bellintani Jr and A.A. Ferreira and A.M.M. Fonseca and I.C. Nascimento and L.F. Ruchko and E.A.O. Saettone and V.S. Tsypin and O.C. Usuriaga},
title = {Runaway discharges in TCABR},
journal = {Nuclear Fusion}
}

@article{bandaru2021magnetohydrodynamic,
  title={Magnetohydrodynamic simulations of runaway electron beam termination in JET},
  author={Bandaru, V. and Hoelzl, M and Reux, C and Ficker, O and Silburn, S and Lehnen, M and Eidietis, N and Contributors, JET and JOREK Team and others},
  journal={Plasma Physics and Controlled Fusion},
  volume={63},
  number={3},
  pages={035024},
  year={2021},
  publisher={IOP Publishing}
}

@article{Strumberger_2017,
doi = {10.1088/0029-5515/57/1/016032},
url = {https://doi.org/10.1088/0029-5515/57/1/016032},
year = {2016},
publisher = {IOP Publishing},
volume = {57},
number = {1},
pages = {016032},
author = {Strumberger, E. and Günter, S.},
title = {CASTOR3D: linear stability studies for 2D and 3D tokamak equilibria},
journal = {Nuclear Fusion}
}

@article{Strumberger_2019,
doi = {10.1088/1741-4326/ab314b},
url = {https://doi.org/10.1088/1741-4326/ab314b},
year = {2019},
publisher = {IOP Publishing},
volume = {59},
number = {10},
pages = {106008},
author = {Strumberger, E. and Günter, S.},
title = {Linear stability studies for a quasi-axisymmetric stellarator configuration including effects of parallel viscosity, plasma flow, and resistive walls},
journal = {Nuclear Fusion}
}

@article{HIRSHMAN_NEMEC,
title = {Three-dimensional free boundary calculations using a spectral Green's function method},
journal = {Computer Physics Communications},
volume = {43},
number = {1},
pages = {143-155},
year = {1986},
issn = {0010-4655},
doi = {https://doi.org/10.1016/0010-4655(86)90058-5},
url = {https://www.sciencedirect.com/science/article/pii/0010465586900585},
author = {S. P. Hirshman and W.I. {van RIJ} and P. Merkel}
}

@article{Schoonheere_2024,
    author = {Schoonheere, N. and Reux, C. and Meister, H. and Beyer, P. and Carvalho, P. and Coffey, I. and Lawson, K. and Puglia, P. and Sheikh, U. and JET Contributors and the EUROfusion Tokamak Exploitation Team},
    title = {Spurious radiated power signal following massive material injections in JET and the effect of neutral gas pressure on resistive bolometers},
    journal = {Review of Scientific Instruments},
    volume = {95},
    number = {12},
    pages = {123509},
    year = {2024},
    issn = {0034-6748},
    doi = {10.1063/5.0224783},
    url = {https://doi.org/10.1063/5.0224783},
    eprint = {https://pubs.aip.org/aip/rsi/article-pdf/doi/10.1063/5.0224783/20295816/123509_1_5.0224783.pdf},
}

@article{Boozer_2016,
    author = {Boozer, A. H. and Punjabi, A.},
    title = {Loss of relativistic electrons when magnetic surfaces are broken},
    journal = {Physics of Plasmas},
    volume = {23},
    number = {10},
    pages = {102513},
    year = {2016},
    issn = {1070-664X},
    doi = {10.1063/1.4966046},
    url = {https://doi.org/10.1063/1.4966046},
    eprint = {https://pubs.aip.org/aip/pop/article-pdf/doi/10.1063/1.4966046/14029268/102513\_1\_online.pdf},
}

\section*{Acknowledgments}
The author would like to thank V. Igochine, J. Herfindal, and H. Zohm for valuable discussions. This work has been carried out within the framework of the EUROfusion Consortium, funded by the European Union via the Euratom Research and Training Programme (Grant Agreement No 101052200 EUROfusion). Views and opinions expressed are however those of the author(s) only and do not necessarily reflect those of the European Union or the European Commission. Neither the European Union nor the European Commission can be held responsible for them. This material is based upon work supported by the U.S. Department of Energy, Office of Science, Office of Fusion Energy Sciences, using the DIII-D National Fusion Facility, a DOE Office of Science user facility, under Award(s) DE-FC02-04ER54698. This report was prepared as an account of work sponsored by an agency of the United States Government. Neither the United States Government nor any agency thereof, nor any of their employees, makes any warranty, express or implied, or assumes any legal liability or responsibility for the accuracy, completeness, or usefulness of any information, apparatus, product, or process disclosed, or represents that its use would not infringe privately owned rights. Reference herein to any specific commercial product, process, or service by trade name, trademark, manufacturer, or otherwise does not necessarily constitute or imply its endorsement, recommendation, or favoring by the United States Government or any agency thereof. The views and opinions of authors expressed herein do not necessarily state or reflect those of the United States Government or any agency thereof. The Swiss contribution to this work has been funded by the Swiss State Secretariat for Education, Research and Innovation (SERI). Large Language Models were used during the preparation of this manuscript to assist with grammar checking and to improve readability. All scientific content, interpretations, and conclusions have been carefully verified by the authors.

\end{document}